\PassOptionsToPackage{table}{xcolor}
\documentclass[acmsmall,screen]{acmart}
\AtBeginDocument{%
  }

\setcopyright{acmlicensed}
\copyrightyear{2026}
\acmYear{2026}
\acmDOI{XXXXXXX.XXXXXXX}

\usepackage{enumitem}
\setlist{leftmargin=*, nosep}

\usepackage{xcolor}
\usepackage{colortbl}
\usepackage{subcaption}
\usepackage{multirow}
\usepackage{tikz}
\usepackage{relsize} 
\usepackage{booktabs}
\usepackage{threeparttable}
\usepackage{siunitx}
\usepackage{tabularx}
\usepackage[utf8]{inputenc}  
\usepackage{listings}
\usepackage[most]{tcolorbox}
\tcbuselibrary{listings,breakable,skins}
\usepackage{xspace}
\usepackage{url}
\usepackage[title]{appendix}%
\hypersetup{hidelinks,breaklinks=true}   

\tcbset{boxrule=1pt, top=2pt, bottom=2pt, left=3pt, right=3pt, fontupper=\small}

\definecolor{rqslate}{RGB}{38,70,108}
\newtcolorbox{rqbox}[1]{%
  enhanced, 
  colback=rqslate!4, colframe=rqslate,
  boxrule=0.6pt, arc=2pt, outer arc=2pt,
  top=9pt, bottom=5pt, left=7pt, right=7pt,
  fontupper=\small,
  fonttitle=\bfseries\small, coltitle=white,
  attach boxed title to top left={xshift=10pt, yshift=-\tcboxedtitleheight/2},
  boxed title style={colback=rqslate, colframe=rqslate, boxrule=0pt,
                     arc=2pt, top=2pt, bottom=2pt, left=6pt, right=6pt},
  title={#1},
}

\DeclareRobustCommand{\circnum}[2][red]{%
  \tikz[baseline=(char.base)]{
    \node[
      shape=circle,
      draw=#1,      
      fill=white,   
      text=#1,      
      line width=0.4pt,
      inner sep=0.1ex,      
      minimum size=1.8ex     
    ] (char) {\scriptsize #2}; 
  }%
}

\newtcblisting{promptbox}[2][]{
  listing only,
  listing options={
    basicstyle=\footnotesize\ttfamily,
    breaklines=true,
    breakatwhitespace=false,
    columns=flexible,
    keepspaces=true,
  },
  colback=gray!5, 
  colframe=black!20, 
  boxrule=0.4pt, 
  arc=2pt,
  left=6pt,
  right=6pt,
  top=6pt,
  bottom=6pt, 
  breakable,
  title={#2},
  label={#1},
}

\newcommand{\appname}{HAFixAgent\xspace}

\newcommand{\nonhistory}{\textit{non-history}\xspace}
\newcommand{\fnall}{\textit{fn\_all}\xspace}
\newcommand{\fnpair}{\textit{fn\_pair}\xspace}
\newcommand{\fldiff}{\textit{fl\_diff}\xspace}

\begin{document}

\title{\appname: History-Aware Agentic Program Repair}


\author{Yu Shi}
\affiliation{%
  \institution{Queen's University}
  \city{Kingston}
  \country{Canada}}
\email{y.shi@queensu.ca}
\orcid{0009-0005-6083-0932}

\author{Hao Li}
\affiliation{%
  \institution{Queen's University}
  \city{Kingston}
  \country{Canada}}
\email{hao.li@queensu.ca}
\orcid{0000-0003-4468-5972}

\author{Bram Adams}
\affiliation{%
  \institution{Queen's University}
  \city{Kingston}
  \country{Canada}}
\email{bram.adams@queensu.ca}
\orcid{0000-0001-7213-4006}

\author{Ahmed E. Hassan}
\affiliation{%
  \institution{Queen's University}
  \city{Kingston}
  \country{Canada}}
\email{ahmed@cs.queensu.ca}
\orcid{0000-0001-7749-5513}

\renewcommand{\shortauthors}{Yu Shi et al.}

\begin{abstract}
  Automated program repair (APR) has recently shifted toward large language models and agent-based systems, with prior work showing that repository history helps repair single-line bugs, since the last commit touching the buggy line is often the bug-introducing one. In this paper, we conduct a large-scale empirical study of whether, how much, and under what conditions repository history improves agentic APR beyond this single-line setting. We present \textbf{\appname}, a history-aware program repair agent that injects blame-derived repository context into its repair loop. After a preliminary study on 854 Defects4J (Java) and 501 BugsInPy (Python) bugs showing that bug-relevant history is widely available across both benchmarks, our evaluation of \appname across three base models (DeepSeek-V3.2-Exp, Qwen3-Coder-Next, and Devstral-Small-2) under perfect fault localization shows that the best history heuristic's improvement over non-history ranges from modest on the strongest model (+4.4\% on Defects4J and +1.1\% on BugsInPy) to large on the weaker models (up to +64.3\%). \appname also outperforms both baselines on the strongest model, RepairAgent by +56.6\% and BIRCH-feedback by +47.1\%, showing it is a strong repair agent independent of history. Furthermore, we observe a strong complementarity between the three heuristics: each repairs a different subset of bugs, so a hybrid approach corresponding to the union of the three heuristics repairs significantly more bugs than non-history (McNemar's test, $p<0.001$). This complementarity even survives when moving from a perfect fault localization setting to a more realistic spectrum-based fault localization. While an individual history heuristic is generally no more expensive than non-history in terms of agent steps or tokens under both perfect and realistic FL, the hybrid approach costs roughly twice a single run when used in a cascade setting, which is still feasible given the improved performance.
\end{abstract}





\maketitle

\newcommand{\rqzero}{To what extent is historical information available in real-world bugs?}
\newcommand{\rqone}{How effective is \appname with perfect fault localization?}
\newcommand{\rqtwo}{How effective is \appname with realistic fault localization?}
\newcommand{\rqthree}{How efficient is \appname?}

\section{Introduction}
Since manually identifying and fixing software bugs consumes significant developer time and resources, automated program repair~(APR) has been an active and promising research direction for over a decade, aiming to mitigate this burden by automatically generating patches for detected bugs~\citep{goues2019apr}.
APR has evolved from heuristic-based \citep{GenProg,le2016history,wen2018context}, constraint-based \citep{le2017s3,mechtaev2016angelix,long2015staged}, and template-based techniques \citep{liu2019tbar,martinez2016astor,liu2019avatar} to learning-based methods~\citep{chen2019sequencer,jiang2021cure,li2020dlfix,lutellier2020coconut,zhu2021syntax,zhu2023tare,ye2024iter} and, most recently, large language model-based approaches \citep{zhang2024systematic,jiang2023impact,hossain2024deep,xia2023automated,xia2024automated,xia2023plastic,fan2023automated,yin2024thinkrepair,huang2025template,li2024hybrid,shi2025hafix} and agent-based approaches \citep{yang2024sweagent,bouzenia2024repairagent,mu2025experepair,meng2024empirical,lee2024unified,zhang2024autocoderover,chen2025swe,wang2024openhands}. However, despite substantial progress, reliably repairing complex bugs remains challenging.

Particularly challenging are bugs that span multiple lines and locations (e.g., multi-hunks), which require coordinated edits across those locations. As \citet{nashid2025characterizing} highlight, these bugs are common yet significantly harder for current LLM-based approaches. To address them, recent APR has turned to agent-based approaches~\citep{bouzenia2024repairagent,yang2024sweagent,wang2024openhands} that let LLMs use tools iteratively over multiple turns (code search, editing, testing). These agents, however, work only from the current code snapshot and ignore repository history. Decades of mining software repositories (MSR) research~\citep{sliwerski2005changes,hassan2006mining,adams2010identifying} show that this history is valuable for understanding and fixing bugs, and prior work~\citep{shi2025hafix} showed that injecting context from git blame\footnote{\url{https://git-scm.com/docs/git-blame}} commits improves LLM repair, but only for single-line bugs.

We hypothesize that adapting history-based context from the single-line setting to multi-line and multi-file bugs, and giving it to an agent that acts on it over multiple turns, can improve their repair as well. History provides information on relevant past changes, developer rationale, and co-evolution patterns, while an agent can exploit this rich history across multiple repair turns. Yet it is unclear whether this benefit materializes at scale, whether it survives realistic fault localization, and whether the three history heuristics are redundant or genuinely complementary.

This paper presents \textbf{\appname}, a history-aware agent for APR that integrates history heuristics (inspired by HAFix~\citep{shi2025hafix}) to enable the agent to leverage historical context alongside bash command tools during its autonomous execution. To evaluate the effectiveness of \appname, we apply it to all 854 bugs in Defects4J~\citep{just2014defects4j} (Java) and all 501 bugs in BugsInPy~\citep{widyasari2020bugsinpy} (Python), two widely used benchmarks for evaluating APR techniques. In this study, we address the following research questions~(RQs):

\smallskip\noindent\textbf{RQ0. \rqzero}
Before integrating repository history, we first establish whether relevant historical information is broadly available for use in APR, since it is not always present: some bugs are fixed only by inserting missing or new code, which \texttt{git blame} cannot annotate, leaving them \emph{blameless}. We analyze both Defects4J and BugsInPy to measure the availability and concentration of historical information. Our findings confirm that 71.1\% (Defects4J) and 87.4\% (BugsInPy) of bugs have available historical information. In Defects4J, 99.5\% of blameable bugs map to a single blame commit, while BugsInPy shows more scattered blame (58.4\% single-commit).

\smallskip\noindent\textbf{RQ1. \rqone}
We compare \appname against two state-of-the-art baselines replicated on the same LLMs, then isolate the effect of history through a controlled comparison (history versus non-history) on Defects4J and BugsInPy. On DeepSeek-V3.2-Exp, \appname outperforms both baselines, RepairAgent by +56.6\% and BIRCH-feedback by +47.1\%. These margins are specific to DeepSeek, since on the two smaller models \appname's best configuration matches or beats the baselines. Under perfect fault localization, the net benefit of history on DeepSeek is modest (+4.4\% on Defects4J, +1.1\% on BugsInPy), but the three history heuristics are complementary. Each repairs a different subset of bugs, so a hybrid approach (the union of the three) repairs 684 of 854 Defects4J and 463 of 501 BugsInPy bugs, versus 522 and 373 for non-history (McNemar's test, $p<0.001$).

\smallskip\noindent\textbf{RQ2. \rqtwo}
Because perfect fault localization is unrealistic, we re-evaluate all configurations under realistic spectrum-based fault localization (SBFL), at full scale across both benchmarks. The complementarity of the history-based heuristics persists: even under noisy localization, the hybrid approach significantly outperforms non-history on both benchmarks (679 resolved bugs versus 466 on Defects4J, 299 versus 209 on BugsInPy).

\smallskip\noindent\textbf{RQ3. \rqthree}
We analyze inference cost and the number of agent reasoning steps across all configurations under both perfect and realistic fault localization. We find that adding historical context does not significantly change inference cost or the number of agent steps relative to non-history, except on single-line Defects4J bugs and for the \fnall heuristic.

\smallskip
The main contributions of our paper are as follows:

\begin{itemize}
  \item The design and implementation of \appname, a novel history-aware agentic approach for APR that enriches its decision-making loop with deep repository context.
  \item A practical and automated method for constructing historical context, featuring three distinct blame-derived historical heuristics and a robust fallback strategy for any agentic APR system.
  \item A large-scale empirical study on 854 Defects4J (Java) and 501 BugsInPy (Python) bugs, with same-LLM baseline replication, providing evidence that history (1) is broadly available across both benchmarks, (2) improves agent effectiveness, where individual heuristics help and their hybrid significantly outperforms non-history under both perfect and realistic (SBFL) fault localization, with the advantage holding under manual patch-correctness inspection, (3) generalizes across base models and especially helps weaker, cheaper ones perform better, and (4) does not increase the per-bug cost of an individual history heuristic over non-history, except on single-line Defects4J bugs and for the \fnall heuristic, though the hybrid costs roughly 1.4 to 2.0 times a single run.
\end{itemize}

\section{Background and Related Work}\label{sec:background}


\subsection{Traditional Automated Program Repair}\label{subsec:traditional-apr}
Traditional APR formulates patch generation as a search or constraint problem over program transformations guided by tests or specifications. Search-based systems evolve patches with genetic or heuristic search~\citep{GenProg,le2016history,wen2018context}. Semantics-based repair solves synthesis constraints~\citep{nguyen2013semfix,mechtaev2016angelix,xuan2016nopol}. Template-based repair uses mined fix schema~\citep{liu2019tbar,liu2019avatar} or in-project code fragments~\citep{DBLP:conf/kbse/XinR17,wen2018context}. Learning-based APR before LLMs frames repair as code-to-code translation, including SequenceR~\citep{chen2019sequencer}, CoCoNuT~\citep{lutellier2020coconut}, CURE~\citep{jiang2021cure}, and Recoder~\citep{zhu2021syntax}, with production systems like Getafix~\citep{bader2019getafix} and SapFix~\citep{marginean2019sapfix}. SelfAPR~\citep{DBLP:conf/kbse/YeML0M22} further advances this line by using self-supervised learning on project-specific perturbations with test execution diagnostics, and RewardRepair~\citep{ye2022neural} incorporates execution-based signals into the training objective to reduce test overfitting. These approaches established core pipelines but are constrained by operator design, search scalability, and overfitting to tests.

\subsection{LLM and Agent-based APR}\label{subsec:LLM-agent-based-APR}
Large language models have significantly advanced APR, outperforming prior learning-based approaches~\citep{jiang2023impact,xia2023automated}. Conversation-driven repair simulates debugging by interleaving generation with test feedback~\citep{xia2024automated}. Self-directed repair uses chain-of-thought to gather knowledge before fixing~\citep{yin2024thinkrepair}. Hybrid and template-guided methods constrain the patch space with analysis or repair templates~\citep{li2024hybrid,huang2025template,xia2023plastic}.

Agent-based APR represents a further step, shifting from single-shot generation to autonomous, tool-using workflows. An LLM-based agent couples reasoning with memory and tools within an iterative execution loop, often conceptualized as a Reason-Act (ReAct) cycle~\citep{yao2023react,wang2024survey}. This approach is central to Agentic Software Engineering~\citep{hassan2024rethinking,hassan2025agentic,li2025rise,roychoudhury2025agentic}. SWE-agent~\citep{yang2024sweagent} introduced an Agent-Computer Interface with developer-centric tools for resolving GitHub issues. OpenHands~\citep{wang2024openhands} builds generalist agents, and RepairAgent~\citep{bouzenia2024repairagent} provided dedicated toolsets for code reading, patch generation, and test feedback. AutoCodeRover~\citep{zhang2024autocoderover} enhanced agents with program analysis context, while multi-agent collaboration~\citep{lee2024unified} and empirical studies~\citep{meng2024empirical} are mapping this design space. A recent trend incorporates memory and experience from prior repair trajectories~\citep{mu2025experepair,chen2025swe}. Prometheus~\citep{pan2025prometheus} further introduces a memory-centric framework with a unified knowledge graph for long-horizon codebase navigation. Agentless~\citep{xia2024agentless} questions whether full autonomy is necessary, showing that a simple localize-repair-validate pipeline can rival agent-based systems. However, these systems generally rely on local code and runtime outputs, with limited study of commit history as a first-class context inside the loop, which is the focus of this work.

\subsection{Context and History for APR}\label{subsec:context-history-apr}
A core question for LLM-based APR is which context to provide. Prior work shows that adding bug-related facts such as error messages, stack traces, and failing tests improves repair~\citep{parasaram2024fact,xia2024automated}, and that models are sensitive to how much code surrounds the edit region~\citep{prenner2024out}. Other work leverages natural language bug reports~\citep{fakhoury2024nl2fix,haque2025towards} or repository-level evaluations~\citep{chen2024large}.

Beyond local signals, decades of Mining Software Repositories (MSR) research have established that historical data is a powerful asset for understanding and fixing bugs~\citep{hassan2006mining,sliwerski2005changes,adams2010identifying,kamei2012large}. \citet{le2016history} pioneered history-driven repair by mining bug-fix patterns that recur across many projects and prioritizing patch candidates that match them. This differs from our history: we retrieve the specific commit that last modified the buggy lines, not fixes to similar bugs elsewhere. More recently, \texttt{git blame}, which annotates each line with the commit that last modified it, provides a direct heuristic for identifying the bug-introducing commit, aligning with the SZZ algorithm's core assumption~\citep{sliwerski2005changes,kim2006automatic}. The last commit to touch a line is, however, a heuristic proxy for, not a guarantee of, the true bug-inducing commit (BIC). A separate line of work identifies BICs more precisely: FONTE ranks commits by fault-localization-weighted suspiciousness from a failure~\citep{DBLP:conf/icse/AnHKY23}, SemBIC statically tracks semantic changes along the failing execution path across history~\citep{DBLP:journals/pacmse/ChenZCWC25}, and LLM4SZZ augments SZZ with LLM-based assessment~\citep{DBLP:journals/pacmse/TangLLYB25}.

\appname adopts the lightweight blame proxy and measures its aggregate effect, and integrating principled BIC identification is a complementary direction. HAFix~\citep{shi2025hafix} demonstrated that injecting this blame-derived context significantly improves repair for single-line bugs. Layered knowledge injection~\citep{ehsani2025bug} systematically adds project and repository knowledge to prompts. \appname advances this direction by bringing blame-driven history into the agentic workflow and quantifying its effect on both single-hunk and multi-hunk bugs across two benchmarks.

\section{Preliminary study (RQ0): \rqzero}\label{sec:preliminary-study}

\subsection{Motivation}
Providing LLM agents with informative context is essential for effective repair \citep{prenner2024out}. While most context engineering approaches for automated program repair (APR) focus on knowledge obtained from a given snapshot of a code base \citep{prenner2024out,haque2025towards,fakhoury2024nl2fix,chen2024large,xia2023automated,li2024hybrid}, only recently APR techniques were proposed that leverage code history \citep{shi2025hafix,ehsani2025bug}. Inspired by the domain of mining software repositories~\citep{sliwerski2005changes,hassan2006mining,kamei2012large}, prior work on single-line APR~\citep{shi2025hafix} uses context related to the last commit that touched the bug location, as the latter commit commonly is assumed to be the bug-introducing commit and hence has valuable knowledge about the introduced bug. Finding this commit for single-line bugs is straightforward (as obtained using the \texttt{git blame} command), as they only comprise one buggy line.

Unfortunately, more complex bugs touching one hunk, multiple hunks in one file or multiple hunks across files potentially would require combining blame commit information of many different lines, potentially drowning the latter information within an LLM's token window, and hence reducing the performance of historical context for APR. In such a case, an APR technique would need to aggregate and rank multiple blame history sources, increasing prompt size, latency, and the risk of distraction.

Furthermore, for a given buggy snapshot of the code base, blame commits do not exist for all buggy lines. For instance, when the fault location is actually an insertion point (where code currently is missing) rather than an existing line that needs to be modified, \texttt{git blame} cannot be applied, as it can only annotate existing code. Hence, the more likely bug fixes insert lines instead of modifying them, the less applicable a blame-based historical APR technique would become, unless a workaround can be found.

This preliminary RQ studies how much of a challenge the scale of blameable history is for single-hunk, multi-hunk and multi-file bugs: (i) how often does a blame commit exist for buggy lines? and (ii) across how many unique commits is the historical blame information of these types of bugs spread? The results will provide clear empirical insights about the effort required to adapt history-based context to more complex types of bugs.

\subsection{Approach}\label{subsec:prelim-setup}

\subsubsection{Dataset}
To study the availability and distribution of historical information in real-world bugs across languages, we choose two established benchmarks.
\textbf{Defects4J} (v3.0.1) \citep{just2014defects4j} is a standard and widely-used benchmark for evaluating recent APR techniques~\citep{xia2022less,jiang2023impact,hossain2024deep,lutellier2020coconut,jiang2021cure,ye2024iter,bouzenia2024repairagent}. It contains 854 Java bugs alongside test cases for each bug collected from 17 different open-source projects, spanning various domains such as data visualization (Chart), compiler construction (Closure), and date/time utilities (Time).
\textbf{BugsInPy}~\citep{widyasari2020bugsinpy} is a benchmark of 501 Python bugs collected from 17 popular open-source projects, including data processing (pandas), web frameworks (fastapi, tornado), and utility libraries (tqdm, thefuck). BugsInPy complements Defects4J by covering a dynamically typed language with different development practices and commit patterns.

In this preliminary RQ, we use the oracle bug fix provided by each benchmark in a post-hoc manner to identify the buggy lines (i.e., the lines in the buggy snapshot that were modified or deleted in the fix), then run \texttt{git blame} on exactly these lines to obtain the most recent commits that previously modified them. We then de-duplicate the commit hashes for each bug, count each bug's unique blame commits, and compute the distribution by bug category.

\subsubsection{Bug Category Terminology}\label{subsubsec:bug-category-terminology}
To categorize bugs based on the structure of their bug fix commits, we first define a \textbf{hunk} as a single, contiguous block of edits (i.e., additions, deletions, or modifications) applied to a specific region in the source code, similar to earlier work~\citep{nashid2025characterizing}. A patch file can contain one or more such hunks, and a bug fix can change more than one such file. Furthermore, following prior work on bug categorization~\citep{nashid2025characterizing,bouzenia2024repairagent,jiang2023impact,xia2024automated}, we classify bugs in both datasets into four mutually exclusive categories based on the number and location of these hunks:
(1) \textbf{Single-Line (SL)}: The fix edits exactly one line in one file. (2) \textbf{Single-Hunk (SH)}: The fix edits one contiguous hunk of multiple lines in one file. (3) \textbf{Single-File-Multi-Hunk (SFMH)}: The fix contains two or more distinct hunks confined to a single file. (4) \textbf{Multi-File-Multi-Hunk (MFMH)}: The fix contains hunks that span two or more different files. Unlike prior work that either nests SL within SH \citep{jiang2023impact} or collapses SFMH and MFMH into one multi-hunk class \citep{nashid2025characterizing}, we keep all four categories separate for finer granularity.

We also categorize bugs by blame availability, i.e., whether the buggy version contains an existing faulty line that fault localization could point to and \texttt{git blame} can annotate. (1) \textbf{Blameable}: the bug fix modifies or deletes at least one existing line in the buggy snapshot, so such a faulty line exists for fault localization to identify and \texttt{git blame} to annotate. The fix is therefore not purely new code. (2) \textbf{Blameless}: the bug fix only inserts new lines or files, with no deletions or modifications, so no existing faulty line exists, and fault localization can at most indicate where new code is needed.


\subsection{Results}
\begin{figure}[t]
  \centering
  \includegraphics[width=\textwidth]{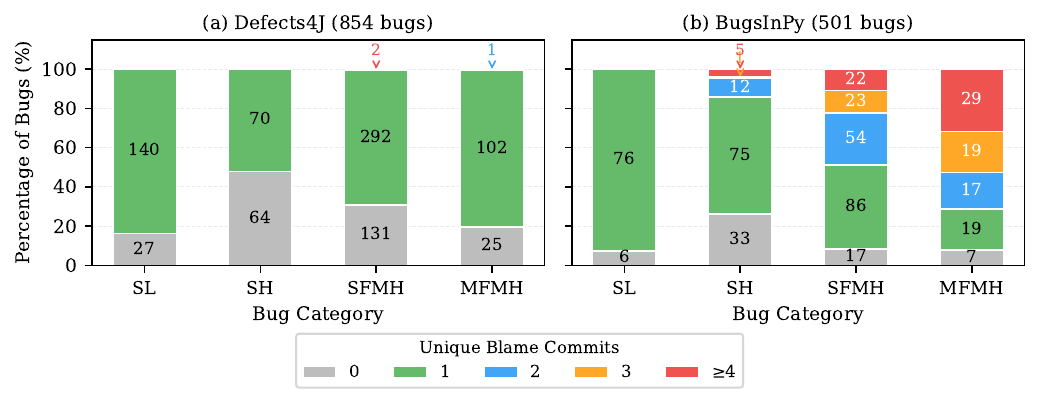}
  \caption{Distribution of the number of unique blame commits per bug category on (a) Defects4J and (b) BugsInPy. Defects4J blame is almost entirely concentrated in a single commit, while BugsInPy is often multi-commit for complex bug categories (49.0\% of SFMH and 71.4\% of MFMH bugs).}
  \label{fig:rq0_blame_commit_distribution}
\end{figure}
\textbf{Historical blame commit information is broadly available, even for complex bugs, with 71.1\% of Defects4J and 87.4\% of BugsInPy bugs blameable.} In Defects4J, complex multi-hunk bugs (SFMH and MFMH), which account for 64.8\% of the dataset, maintain high history availability: 69.2\% of SFMH and 80.5\% of MFMH bugs are blameable. SH has the lowest rate at 52.2\%, as single-hunk bugs are more frequently pure insertions. In BugsInPy, blame availability is even higher across all categories, with SL at 92.7\%, SFMH at 91.6\%, and MFMH at 92.3\%, while SH remains the lowest at 73.8\%, consistent with the Defects4J pattern.
These results indicate that historical blame data is commonly available across bug categories and both datasets, supporting history-aware APR techniques. For bugs with no blame commit (28.9\% in Defects4J, 12.6\% in BugsInPy), a workaround would be required.

\textbf{Blame concentration differs sharply across datasets: Defects4J is usually single-commit, whereas BugsInPy is often multi-commit.} Figure~\ref{fig:rq0_blame_commit_distribution}(a) shows that 70.7\% (604 out of 854) of Defects4J bugs map to exactly one unique blame commit, and only 0.4\% (3 out of 854) have two or more. Manual inspection confirmed that multi-hunk bugs typically trace back to a single common commit, as their buggy lines largely overlap in history. In contrast, BugsInPy blame is more scattered, with 51.1\% (256 out of 501) of bugs mapping to a single blame commit and 36.3\% (182 out of 501) having two or more. The multi-commit proportion rises with bug complexity: 0\% for SL, 14.3\% for SH, 49.0\% for SFMH, and 71.4\% for MFMH. Among blameable bugs, 58.4\% in BugsInPy are single-commit, compared to 99.5\% in Defects4J. This contrast likely reflects development practices. In BugsInPy, the buggy lines of a single bug were often last modified by several different commits, so \texttt{git blame} maps that bug to multiple commits. In Defects4J, those lines were more often introduced together by one commit, so a single blame commit covers them.

\begin{rqbox}{Summary for RQ0}
  \begin{enumerate}
    \item Blame commits are available for the majority of bugs in both benchmarks: 71.1\% in Defects4J (Java) and 87.4\% in BugsInPy (Python).
    \item Blame concentration differs across datasets: 99.5\% of blameable Defects4J bugs map to a single commit, whereas this drops to 58.4\% in BugsInPy, where multi-commit blame increases with bug complexity.
  \end{enumerate}
\end{rqbox}

\section{\appname}\label{sec:HAFixAgent}

Based on the findings of the preliminary study, we propose a new APR approach called \appname. Prior history-aware repair~\citep{shi2025hafix} feeds blame context into a single static prompt effective only for single-line fixes, so \appname pairs this context with an agent for the iterative, multi-location and perhaps multi-step repair that single-line prompting cannot do. It exhibits the following three design decisions:
\begin{enumerate}
  \item a minimal agent-based architecture designed to iteratively handle complex, multi-step reasoning and actions (e.g., code exploration, patch validation) required for bugs that go beyond simple, single-line fixes.
  \item the agent has access to a git blame tool, inspired by the findings in Section~\ref{sec:preliminary-study}, and leverages the blame commit as the primary source of historical context. When multiple blame commits exist, an LLM-based judge selects the most relevant one.
  \item the agent includes a fallback mechanism to obtain the relevant history for blameless bugs where no buggy line has an associated blame commit.
\end{enumerate}

Unlike agent-based APR systems such as \textit{RepairAgent} \citep{bouzenia2024repairagent}, which employ pre-defined Python tool APIs, \appname is intentionally minimal and history-aware. It treats repository history from previous snapshots as an important source of context that augments the current code snapshot, and it operates within a compact, guarded loop over a small set of standard bash tools (e.g., \texttt{grep}, \texttt{sed}, \texttt{find}). This design choice is critical: it minimizes ambiguity from complex and high-level tool interactions, allowing us to more directly attribute performance gains to the quality of the historical context rather than to sophisticated tool exploration. Additionally, the design is consistent with recent commercial coding assistants such as Claude Code \citep{anthropic_claude_code_docs_2025} and OpenAI Codex \citep{openai_codex_2025}.

\begin{figure}[t]
  \centering
  \includegraphics[width=0.7\textwidth]{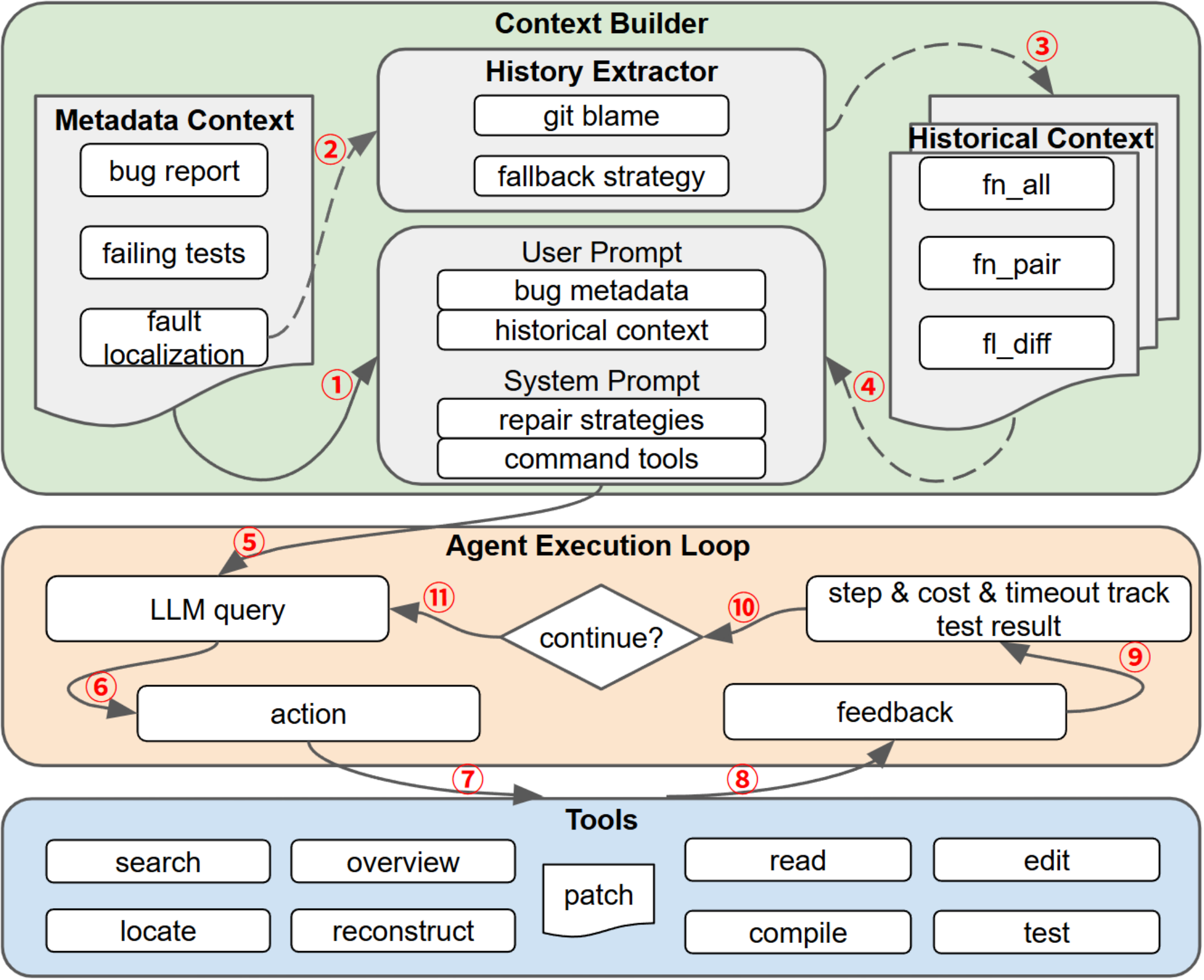}
  \caption{\appname architecture and workflow.}
  \label{fig:HAFixAgent_Architecture}
\end{figure}

Figure~\ref{fig:HAFixAgent_Architecture} overviews the architecture. The \textbf{Context Builder} assembles metadata context \circnum{1} and uses the History Extractor to retrieve \circnum{2} and format \circnum{3} historical context into the prompt \circnum{4}. The \textbf{Agent Execution Loop} drives iterative repair: the LLM generates an action \circnum{5}\circnum{6}, which executes in the \textbf{Tools} sandbox \circnum{7}. Feedback \circnum{8} is parsed for test results \circnum{9}, checked against termination guards \circnum{10}, and fed back to the LLM \circnum{11}.

\subsection{Context Builder}\label{subsec:context-builder}

\subsubsection{Metadata Context}\label{subsubsec:metadata-context}
The user prompt contains three items: the human-written bug report, the initial set of failing tests, and the fault localization context, which specifies the buggy files and lines. These metadata are used only as \nonhistory context and are rendered in step \circnum{1}.

\subsubsection{History Extractor}\label{subsubsec:historical-extractor}

The \textit{History Extractor} module (step \circnum{2} in Figure~\ref{fig:HAFixAgent_Architecture}) finds a relevant blame commit for every bug. We use the buggy lines identified through the fault localization metadata to obtain the blame commit, since the historical context in that blame commit may reveal the root causes of the bug \citep{sliwerski2005changes,hassan2006mining,kamei2012large}. The module handles blameable and blameless bugs differently:

\begin{itemize}
  \item \textbf{Blameable bugs}. For bugs involving modified or deleted lines, we execute \texttt{git blame} on all identified lines and collect the union of all unique commit hashes. As demonstrated in our preliminary study (Section~\ref{sec:preliminary-study}), this union is often a single commit (99.5\% of blameable Defects4J bugs), though BugsInPy bugs more frequently have multiple blame commits (41.6\%). When multiple commits exist, we use an LLM-as-a-judge to select the most relevant one (detailed in Section~\ref{subsubsec:data-collection}).

  \item \textbf{Blameless bugs} (fallback strategy). To enable these blameless bugs (i.e., add-only fixes) to also benefit from historical context, we design a fallback strategy. Specifically, we employ a fallback strategy to blame the nearest executable code line (e.g., an executable line rather than a comment or symbol) within the five lines preceding the insertion. This strategy is based on the assumption that new code is often added near existing, related logic, making the most recent commit to that local area a relevant source of historical context.
\end{itemize}

\subsubsection{Historical Context}\label{subsubsec:historical-context}

Once the blame commits are determined by the \textit{History Extractor}, step \circnum{3} then mines relevant historical context from the identified blame commits by leveraging one of the three top-performing historical heuristics from the HAFix approach~\citep{shi2025hafix} (which focused on SL bugs only):

\begin{itemize}
  \item All functions' names from the blame commit (\textbf{\fnall}): includes the names of all functions from all co-changed files in the blame commit. This setting captures a broad structural view of the co-evolved functions within the same history commit.
  \item Historical function body before and after (\textbf{\fnpair}): which contains the before and after snapshot of the function code body that includes the buggy line. This historical context helps the agent understand how the buggy function evolved and provides possible clues about the root cause of the bug.
  \item File-level diff (\textbf{\fldiff}): includes the diff patch obtained from the git diff command in the blame commit. This configuration exposes the exact code modification across files, allowing the agent to reason about fine-grained edits that may have caused the bug.
\end{itemize}

In step \circnum{4}, \appname integrates the history context, which was extracted in the previous step \circnum{3} using one of the three heuristics from the blamed commit. The history is injected directly into the User Prompt (as shown in Figure~\ref{fig:HAFixAgent_Architecture}), with truncation safeguards (e.g., truncating an exceptionally large file diff) applied to ensure that the total prompt fits within the LLM's context window. This process leads to our four context configurations: the agent operates either with the \nonhistory{} context only (from Section \ref{subsubsec:metadata-context}) or with the \nonhistory{} context augmented by one of the three history heuristics (\fnall, \fnpair, or \fldiff).

\subsection{Agent Execution Loop}\label{subsec:agent-execution-loop}
\appname's execution loop follows an observe-act-feedback paradigm, inspired by frameworks like ReAct~\citep{yao2023react} and those used in recent software engineering agents~\citep{yang2024sweagent}. The loop proceeds with the following steps. \circnum{5} The LLM receives the current state (including the full conversation history of prior actions and observations). \circnum{6} The agent parses the model output to extract a tool invocation (i.e., a bash command, which the system prompt constrains the model always to provide) as the next action. \circnum{7} The command runs in a containerized tool sandbox that exposes repair utilities and dataset tools (e.g., search, edit and compile). \circnum{8} Stdout and stderr are collected from the output of command execution as feedback. \circnum{9} Compilation and test outcomes are captured and parsed. \circnum{10} The loop terminates on success or when a termination guard (e.g., step limit, cost limit, or timeout) is reached. \circnum{11} Otherwise, the observation is appended to the conversation history and fed back to the LLM for the next reasoning step.

\subsection{Tools}\label{subsec:tools}

The agent invokes standard bash tools (\texttt{grep}, \texttt{sed}, \texttt{find}, and \texttt{head}/\texttt{tail} for file operations, \texttt{compile} and \texttt{test} for project validation, and \texttt{echo} for completion signaling), with optional chaining via \texttt{\&\&} or \texttt{||}. These actions are executed between steps \circnum{7} and \circnum{8} in the loop, and the full toolset is embedded in the system prompt (available in the replication package).

\appname intentionally exposes a simple but powerful set of bash tools together with project commands like \texttt{compile} and \texttt{test}. A key design choice is that the historical context is not an interactive tool, but is instead modularized and prepared by the \textit{History Extractor} and injected directly into the agent's context before the execution loop begins. This decouples history retrieval from agent actions, which allows the agent's toolset to remain simple, modular, scalable and focused on repair-related operations (e.g., read, edit and test).

In practice, the loop alternates between reading around fault locations, applying edits, compiling, and testing until all tests pass. Termination occurs when the agent emits a completion signal after passing tests, or when a guard (step, cost, or timeout) is reached.

\section{Experimental Setup}\label{subsec:case-study-design}

\subsection{Prompt Context Preparation}\label{subsubsec:data-collection}

Consistent with the preliminary study, we employ the entire 854 bugs of Defects4J and 501 bugs of BugsInPy to prepare the data needed for evaluation.

\textbf{Metadata collection.} To prepare the \nonhistory context as defined in Section~\ref{subsubsec:metadata-context}, we need bug metadata such as bug reports, failing tests, and fault locations. For Defects4J, we mine bug reports from issue tracker links provided by the benchmark. For 18 cases from the \textit{Chart} project without a clear bug issue link, we use the fixing commit message as the bug description, and to prevent potential data leakage, we manually verified these commit messages to ensure they do not contain explicit fix instructions. For BugsInPy, we intentionally omit bug descriptions to avoid data leakage from commit messages, relying on fault locations and failing tests only. Failing tests are collected from each benchmark's test framework. Fault localization is performed as a separate step, described in Section~\ref{subsec:fault-localization}.

\textbf{Historical data collection.} Motivated by the preliminary study in Section~\ref{sec:preliminary-study}, which shows that blame commits are broadly available in both datasets, we collect history per bug as follows:
\begin{itemize}
  \item Blameable with one unique commit: Run \texttt{git blame} on deleted or modified lines to obtain the blame commit, as defined in Section~\ref{subsubsec:historical-context}.
  \item Blameable with multiple commits: In Defects4J, only three bugs fall in this case. In BugsInPy, 182 bugs have two or more blame commits. We use the LLM-as-a-judge (the same LLM as \appname) to select the single most relevant commit. The model is prompted with the set of blameable lines and asked to choose the line which is most likely related to the bug's root cause. The exact judge prompt is included in the replication package.
  \item Blameless with zero commits: Apply the nearest line fallback strategy from Section~\ref{subsubsec:historical-extractor} to select a blame commit.
\end{itemize}

Given the selected blame commit, we extract and construct data of \fnall, \fnpair, and \fldiff as defined in Section~\ref{subsubsec:historical-context}.

\subsection{Baselines}\label{subsec:baselines}

We compare \appname with two recent APR techniques as baselines, i.e., RepairAgent \citep{bouzenia2024repairagent} and BIRCH-feedback \citep{nashid2025characterizing}.

\textbf{RepairAgent} is an autonomous LLM-based agent equipped with pre-defined API tooling. The original study used GPT-3.5-turbo and evaluated on 835 bugs (from Defects4J v1.2 and v2), reporting 186 plausible and 164 correct fixes.
We replicate RepairAgent and evaluate on all 854 Defects4J bugs, using the plausible metric for a direct, same-metric comparison.

\textbf{BIRCH-feedback} targets multi-hunk repairs. The original study used o4-mini and reported results on 371 multi-hunk bugs.
We replicate their best-performing configuration with feedback and evaluate on the same 371 overlapping multi-hunk bugs (244 SFMH + 127 MFMH).

To eliminate the model confound and ensure a fair comparison, we run baselines and \appname with the same three models: DeepSeek-V3.2-Exp~\citep{liu2024deepseek,deepseek-v3.2-exp-tech-note}, Qwen3-Coder-Next~\citep{qwen_qwen3_coder_next_tech_report}, and Devstral-Small-2~\citep{mistral_devstral_2025}.

\subsection{Fault Localization}\label{subsec:fault-localization}

Fault localization~(FL) determines which lines the agent is asked to repair and hence, in the case of all four \appname configurations, which lines \texttt{git blame} runs on to retrieve history. Following common practice in the field of program repair~\citep{yang2024sweagent,fan2023automated,bouzenia2024repairagent,xia2022less,ye2024iter,nashid2025characterizing,ehsani2025bug}, we assume \textit{perfect fault localization} as the default evaluation setting in RQ1: the agent is given the buggy lines, i.e., the lines modified or deleted by the oracle bug fix, extracted following the process in Section~\ref{subsec:prelim-setup}. This depends only on the \textit{location} of the buggy lines, not on the content of the developer's fix, which is never shown to the agent, so it isolates the effect of historical context from fault-localization error. Perfect fault localization is the evaluation setting for RQ1 and RQ3.

Because perfect fault localization is optimistic, RQ2 (Section~\ref{subsec:RQ2}) relaxes this assumption and re-runs all four configurations under \textit{realistic spectrum-based fault localization} (SBFL), which ranks a program's source lines by suspiciousness from test coverage. We use one standard SBFL tool per language, GZoltar~1.7.3~\citep{DBLP:conf/taicpart/RiboiraA10} for Java and FauxPy~0.7.0~\citep{DBLP:journals/corr/abs-2404-18596} for Python, both with the Ochiai formula at line granularity, and take the top-10 ranked lines as the fault locations. These lines replace the oracle patch locations of the perfect fault localization approach, then the history configurations derive their single blame commit from them through the same History Extractor used under perfect FL (Section~\ref{subsubsec:historical-extractor}). In other words, only the localization source changes. The developer's golden patch is used only post hoc, to score repair success and localization accuracy. Section~\ref{subsec:RQ2} reports how noisy this localization is and whether the history benefit survives it.

\subsection{Metrics}\label{subsec:Metrics}

In line with previous studies \citep{chen2021evaluating,parasaram2024fact,ehsani2025bug,nashid2025characterizing,silva2025repairbench}, we report \textit{Plausible@1} to evaluate the correctness based on whether the generated patch passes all test cases. Formally:
\begin{equation}
  \label{eq:plausible1}
  \mathrm{Plausible@1}
  = \frac{1}{|\mathcal{B}|}\sum_{b \in \mathcal{B}}
  \mathrm{TestPass}\!\left(\text{patch}^{\,b}_{1}\right)
\end{equation}
where $\mathcal{B}$ denotes the set of bugs and $\mathrm{TestPass}(\cdot)\in\{0,1\}$ returns $1$ if the patch compiles and all tests pass, and $0$ otherwise. We generate one patch per bug using standard decoding settings, with the temperature set to 0.0 for enhancing reproducibility.

Complementing \textit{Plausible@1}, which measures the pass rate as a proportion of all bugs, we also report \textit{\#Pass} (the absolute count of bugs passing the test suite), similar to prior work \citep{nashid2025characterizing}. To quantify the unique contribution of diverse historical context, we additionally report \textit{\#Unique Pass}, to show the number of bugs solved by a configuration using historical context but not by its \nonhistory counterpart. We visualize overlaps using Venn diagrams to illustrate multi-way complementarity. To test whether one configuration repairs significantly more bugs than another, we use McNemar's exact test, the standard paired test for a binary outcome on the same bugs, treating each bug as repaired or not by both configurations. When comparing the three individual history heuristics and a hybrid approach comprising their union against \nonhistory, we apply a Bonferroni correction for the multiple comparisons.

For efficiency, we report the median number of agent steps per bug, the median cost per bug in USD computed from token usage, and the cost stratified by outcome (successful versus failed fixes). To assess the statistical significance of differences in cost and number of steps, we compare the distributions across the four context configurations. Since the data is paired (same bugs, different configurations) and not assumed to be normally distributed, we employ the Friedman test \citep{friedman1937use}, a non-parametric test for multiple comparisons. If the Friedman test is significant ($p < 0.05$), we conduct pairwise Wilcoxon signed-rank tests \citep{wilcoxon1992individual} for post-hoc analysis to compare each history-aware configuration against the \nonhistory configuration. To control for multiple comparisons (three pairs), we apply the Bonferroni correction \citep{dunn1961multiple} and set the significance threshold $\alpha$.

\subsection{Implementation}\label{subsec:Implementation}

\appname is implemented in Python and builds on \textit{mini-swe-agent} \citep{yang2024sweagent} for the execution loop, which aligns with our lightweight design. While mini-swe-agent (and swe-agent) is designed naturally to support solving PR issue benchmarks such as SWE-Bench \citep{jimenez2023swebench}, it lacks the capability of historical context management (e.g., history localizing and retrieval), the repairing strategies and tools specifically for resolving Defects4J bugs. It is an open-source, lightweight framework with about 100 lines of core loop code, enabling integration of historical context extraction and bash tool orchestration at runtime. The main comparison and history configurations run on DeepSeek-V3.2-Exp via OpenRouter. To test whether the findings hold beyond a single model, we replicate all four configurations and both baselines (RepairAgent and BIRCH-feedback) on two smaller models, Qwen3-Coder-Next and Devstral-Small-2. All runs use temperature 0.0, and prompts are rendered at runtime with \textit{jinja2}.

We run \appname on Ubuntu 20.04, with each bug executing in an isolated Docker container. For Defects4J, the sandbox is built from the released Defects4J image, with the \texttt{compile} and \texttt{test} tools (Section~\ref{subsec:tools}) wrapping \texttt{defects4j compile} and \texttt{defects4j test -r}. For BugsInPy, we build pre-warmed Docker images per project with the correct Python environment, and the \texttt{test} tool wraps \texttt{bugsinpy-test}. For each bug, we cap the agent's loop at 50 steps, enforce a per-bug cost guard of \$1~USD, and set a 1-hour timeout. After each run, the container is automatically cleaned, while all generated patches and logs are persisted for analysis.

\section{Results}\label{sec:evaluation-results}

\subsection{RQ1: \rqone}\label{subsec:RQ1}

\subsubsection{Motivation}\label{subsubsec:RQ1_Motivation}

Prior work~\citep{shi2025hafix} shows that injecting historical context improves LLM repair for single-line bugs under simple prompting. History carries signals that are hard to infer from a static snapshot: the fault-introducing change, coevolving functions and files, and intent hints in commit diffs. Since the preliminary study (Section~\ref{sec:preliminary-study}) shows that this history is broadly available across both benchmarks, this RQ studies whether history improves agent-based repair, which bug types benefit, and whether the benefit holds across base models.

\subsubsection{Approach}\label{subsubsec:RQ1_Approach}

As explained in Section~\ref{subsec:baselines}, we replicate the two state-of-the-art baselines on the same LLMs as \appname to ensure a fair comparison. We compare \appname against \textit{RepairAgent}~\citep{bouzenia2024repairagent} on all 854 Defects4J bugs, and against \textit{BIRCH-feedback}~\citep{nashid2025characterizing} on the same 371 multi-hunk bugs (244~SFMH + 127~MFMH) evaluated in its original study. Both baselines are Defects4J-specific Java systems with no Python configuration, so we compare against them on Defects4J only. The comparison of \appname's history and non-history variants, by contrast, runs on both benchmarks.
To analyze where history helps and where it does not, we categorize results by our four bug types (SL, SH, SFMH, MFMH) and examine each category separately.

In addition, we evaluate the four context configurations of \appname defined in Section~\ref{subsec:context-builder} across all 854 Defects4J bugs and all 501 BugsInPy bugs, i.e., \nonhistory, \fnall, \fnpair, and \fldiff. Blameless bugs (28.9\% in Defects4J, 12.6\% in BugsInPy) use the fallback strategy (Section~\ref{subsubsec:historical-extractor}), ensuring every bug has meaningful historical context. To evaluate whether the benefit is specific to DeepSeek, we replicate the same history-versus-non-history comparison on two additional, code-specialized base models that differ in architecture and scale: Qwen3-Coder-Next~\citep{qwen_qwen3_coder_next_tech_report}, a Mixture-of-Experts model with 80B total and 3B active parameters, and Devstral-Small-2~\citep{mistral_devstral_2025}, a dense 24B model. Both are smaller than DeepSeek-V3.2-Exp and have lower repair rates in our experiments, allowing us to examine how the history benefit varies with baseline repair performance.

Within each model, the four context configurations differ only in the injected historical context. They share the same decoding settings (temperature 0), agent configuration, and containerized runtime. We report both Plausible@1 (computed via termination checks in Section~\ref{subsec:agent-execution-loop}) and \#Pass (the total number of bugs with patches passing all tests). To quantify the specific benefit of historical context, we also report \#Unique Pass for the number of bugs repaired by a history-aware configuration that were not repaired by the \nonhistory configuration.

Finally, all results in this RQ assume \emph{perfect fault localization (FL)}: the ground-truth buggy lines from the developer fix are supplied to the agent as fault locations. This isolates the effect of historical context from localization noise, which RQ2 (Section~\ref{subsec:RQ2}) later relaxes to realistic spectrum-based fault localization.

\subsubsection{Results}\label{subsubsec:RQ1_Results}

\begin{table}[p]
  \centering
  \caption{History versus non-history across the three base models under perfect FL, with the external baselines. Each cell is \#Pass, i.e., the number of bugs plausibly repaired (the Plausible@1 rate is \#Pass$/$Total). \textbf{Hybrid} $=$ number of bugs fixed by at least one of the three history heuristics (\fnall, \fnpair, \fldiff). \textbf{Baseline} $=$ RepairAgent on the full Defects4J set and BIRCH-feedback on the Defects4J multi-hunk subset (its own denominator). Both are Java-only, so it is empty for the BugsInPy rows. Superscripts count the bugs a history configuration repairs that \nonhistory does not. \textbf{Bold} $=$ highest \#Pass per category among the four \appname configurations and the Baseline. Hybrid is an oracle upper bound and is not bolded. A $*$ marks a configuration, an individual history heuristic or their Hybrid (union), that significantly repairs bugs \nonhistory cannot fix (McNemar's exact test versus \nonhistory, Bonferroni $\alpha = 0.05/4 = 0.0125$ across the three heuristics and their union). A dash (-) marks a cell that does not apply.}
  \label{tab:rq2-ablation}
  \scriptsize
  \setlength{\tabcolsep}{3pt}
  \renewcommand{\arraystretch}{0.70}

  \begin{subtable}[t]{\linewidth}
    \centering
    \caption{DeepSeek-V3.2-Exp}
    \label{tab:rq2-ablation-deepseek}
    \resizebox{0.78\textwidth}{!}{%
      \begin{tabular}{l l r r r r r r r}
        \toprule
        \textbf{Dataset} & \textbf{Category} & \textbf{Total} & \textbf{\nonhistory} & \textbf{\fnall}                & \textbf{\fnpair}               & \textbf{\fldiff}                & \textbf{Hybrid} & \textbf{Baseline} \\
        \midrule
        \multirow{5}{*}{\shortstack[l]{Defects4J                                                                                                                                                                               \\(vs.\\RepairAgent)}}
                         & SL                & 167            & 137                  & 136\,$^{\smash{+10}}$          & 136\,$^{\smash{+10}}$          & \textbf{140}\,$^{\smash{+11}}$  & 156\,$^{*}$     & 121               \\
                         & SH                & 134            & 97                   & 96\,$^{\smash{+16}}$           & \textbf{108}\,$^{\smash{+19}}$ & 103\,$^{\smash{+18}}$           & 122\,$^{*}$     & 68                \\
                         & SFMH              & 425            & 238                  & 251\,$^{\smash{+67}}$          & 249\,$^{\smash{+69}}$          & \textbf{254}\,$^{\smash{+68}}$  & 340\,$^{*}$     & 141               \\
                         & MFMH              & 128            & \textbf{50}          & 47\,$^{\smash{+13}}$           & 43\,$^{\smash{+13}}$           & 48\,$^{\smash{+12}}$            & 66\,$^{*}$      & 18                \\
        \cmidrule{2-9}
                         & \textbf{Total}    & 854            & 522                  & 530\,$^{\smash{+106}}$         & 536\,$^{\smash{+111}}$         & \textbf{545}\,$^{\smash{+109}}$ & 684\,$^{*}$     & 348               \\
        \midrule
        \multirow{5}{*}{BugsInPy}
                         & SL                & 82             & \textbf{74}          & 73\,$^{\smash{+4}}$            & 73\,$^{\smash{+7}}$            & 71\,$^{\smash{+5}}$             & 80              & -                 \\
                         & SH                & 126            & 96                   & 100\,$^{\smash{+17}}$          & \textbf{105}\,$^{\smash{+18}}$ & 93\,$^{\smash{+13}}$            & 120\,$^{*}$     & -                 \\
                         & SFMH              & 202            & 146                  & 148\,$^{\smash{+27}}$          & 137\,$^{\smash{+22}}$          & \textbf{157}\,$^{\smash{+31}}$  & 187\,$^{*}$     & -                 \\
                         & MFMH              & 91             & 57                   & 56\,$^{\smash{+19}}$           & \textbf{59}\,$^{\smash{+12}}$  & 53\,$^{\smash{+17}}$            & 76\,$^{*}$      & -                 \\
        \cmidrule{2-9}
                         & \textbf{Total}    & 501            & 373                  & \textbf{377}\,$^{\smash{+67}}$ & 374\,$^{\smash{+59}}$          & 374\,$^{\smash{+66}}$           & 463\,$^{*}$     & -                 \\
        \midrule
        \multirow{3}{*}{\shortstack[l]{Defects4J                                                                                                                                                                               \\multi-hunk\\(vs. BIRCH)}}
                         & SFMH              & 244            & 123                  & 125                            & \textbf{133}                   & 126                             & -               & 85                \\
                         & MFMH              & 127            & \textbf{49}          & 46                             & 42                             & 47                              & -               & 34                \\
        \cmidrule{2-9}
                         & \textbf{Total}    & 371            & 172                  & 171                            & \textbf{175}                   & 173                             & -               & 119               \\
        \bottomrule
      \end{tabular}}
  \end{subtable}

  \vspace{0.02em}

  \begin{subtable}[t]{\linewidth}
    \centering
    \caption{Qwen3-Coder-Next}
    \label{tab:rq2-ablation-qwen}
    \resizebox{0.78\textwidth}{!}{%
      \begin{tabular}{l l r r r r r r r}
        \toprule
        \textbf{Dataset} & \textbf{Category} & \textbf{Total} & \textbf{\nonhistory} & \textbf{\fnall}          & \textbf{\fnpair}                   & \textbf{\fldiff}                  & \textbf{Hybrid} & \textbf{Baseline} \\
        \midrule
        \multirow{5}{*}{\shortstack[l]{Defects4J                                                                                                                                                                               \\(vs.\\RepairAgent)}}
                         & SL                & 167            & 44                   & 55\,$^{\smash{+25}}$     & 67\,$^{\smash{+34\,*}}$            & 65\,$^{\smash{+28\,*}}$           & 95\,$^{*}$      & \textbf{99}       \\
                         & SH                & 134            & 14                   & 24\,$^{\smash{+15}}$     & \textbf{39}\,$^{\smash{+26\,*}}$   & 31\,$^{\smash{+19\,*}}$           & 57\,$^{*}$      & 36                \\
                         & SFMH              & 425            & 60                   & 84\,$^{\smash{+51\,*}}$  & \textbf{95}\,$^{\smash{+61\,*}}$   & 89\,$^{\smash{+57\,*}}$           & 157\,$^{*}$     & 65                \\
                         & MFMH              & 128            & \textbf{11}          & 8\,$^{\smash{+3}}$       & \textbf{11}\,$^{\smash{+6}}$       & 9\,$^{\smash{+4}}$                & 19              & 7                 \\
        \cmidrule{2-9}
                         & \textbf{Total}    & 854            & 129                  & 171\,$^{\smash{+94\,*}}$ & \textbf{212}\,$^{\smash{+127\,*}}$ & 194\,$^{\smash{+108\,*}}$         & 328\,$^{*}$     & 207               \\
        \midrule
        \multirow{5}{*}{BugsInPy}
                         & SL                & 82             & 27                   & 29\,$^{\smash{+8}}$      & 34\,$^{\smash{+14}}$               & \textbf{36}\,$^{\smash{+14}}$     & 54\,$^{*}$      & -                 \\
                         & SH                & 126            & 32                   & 29\,$^{\smash{+12}}$     & 31\,$^{\smash{+14}}$               & \textbf{36}\,$^{\smash{+20}}$     & 56\,$^{*}$      & -                 \\
                         & SFMH              & 202            & 29                   & 32\,$^{\smash{+17}}$     & 40\,$^{\smash{+23}}$               & \textbf{43}\,$^{\smash{+25}}$     & 69\,$^{*}$      & -                 \\
                         & MFMH              & 91             & 3                    & 5\,$^{\smash{+5}}$       & \textbf{7}\,$^{\smash{+6}}$        & 5\,$^{\smash{+5}}$                & 13              & -                 \\
        \cmidrule{2-9}
                         & \textbf{Total}    & 501            & 91                   & 95\,$^{\smash{+42}}$     & 112\,$^{\smash{+57}}$              & \textbf{120}\,$^{\smash{+64\,*}}$ & 192\,$^{*}$     & -                 \\
        \midrule
        \multirow{3}{*}{\shortstack[l]{Defects4J                                                                                                                                                                               \\multi-hunk\\(vs. BIRCH)}}
                         & SFMH              & 238            & 30                   & 36                       & 43                                 & 36                                & -               & \textbf{45}       \\
                         & MFMH              & 120            & 11                   & 7                        & 11                                 & 9                                 & -               & \textbf{15}       \\
        \cmidrule{2-9}
                         & \textbf{Total}    & 358            & 41                   & 43                       & 54                                 & 45                                & -               & \textbf{60}       \\
        \bottomrule
      \end{tabular}}
  \end{subtable}

  \vspace{0.02em}

  \begin{subtable}[t]{\linewidth}
    \centering
    \caption{Devstral-Small-2}
    \label{tab:rq2-ablation-devstral}
    \resizebox{0.78\textwidth}{!}{%
      \begin{tabular}{l l r r r r r r r}
        \toprule
        \textbf{Dataset} & \textbf{Category} & \textbf{Total} & \textbf{\nonhistory} & \textbf{\fnall}               & \textbf{\fnpair}             & \textbf{\fldiff}                   & \textbf{Hybrid} & \textbf{Baseline} \\
        \midrule
        \multirow{5}{*}{\shortstack[l]{Defects4J                                                                                                                                                                               \\(vs.\\RepairAgent)}}
                         & SL                & 167            & 94                   & 88\,$^{\smash{+25}}$          & 97\,$^{\smash{+26}}$         & \textbf{114}\,$^{\smash{+39\,*}}$  & 135\,$^{*}$     & 108               \\
                         & SH                & 134            & 33                   & 59\,$^{\smash{+33\,*}}$       & 53\,$^{\smash{+30\,*}}$      & \textbf{67}\,$^{\smash{+42\,*}}$   & 91\,$^{*}$      & 51                \\
                         & SFMH              & 425            & 76                   & 116\,$^{\smash{+63\,*}}$      & 109\,$^{\smash{+64\,*}}$     & \textbf{132}\,$^{\smash{+79\,*}}$  & 204\,$^{*}$     & 78                \\
                         & MFMH              & 128            & 15                   & 18\,$^{\smash{+10}}$          & \textbf{19}\,$^{\smash{+9}}$ & 17\,$^{\smash{+9}}$                & 31\,$^{*}$      & 15                \\
        \cmidrule{2-9}
                         & \textbf{Total}    & 854            & 218                  & 281\,$^{\smash{+131\,*}}$     & 278\,$^{\smash{+129\,*}}$    & \textbf{330}\,$^{\smash{+169\,*}}$ & 461\,$^{*}$     & 252               \\
        \midrule
        \multirow{5}{*}{BugsInPy}
                         & SL                & 82             & 33                   & 45\,$^{\smash{+20}}$          & 41\,$^{\smash{+17}}$         & \textbf{48}\,$^{\smash{+24}}$      & 66\,$^{*}$      & -                 \\
                         & SH                & 126            & 45                   & 61\,$^{\smash{+29}}$          & 58\,$^{\smash{+28}}$         & \textbf{63}\,$^{\smash{+30\,*}}$   & 87\,$^{*}$      & -                 \\
                         & SFMH              & 202            & 64                   & \textbf{82}\,$^{\smash{+33}}$ & 80\,$^{\smash{+37}}$         & 73\,$^{\smash{+29}}$               & 125\,$^{*}$     & -                 \\
                         & MFMH              & 91             & 18                   & 20\,$^{\smash{+12}}$          & 29\,$^{\smash{+19}}$         & \textbf{33}\,$^{\smash{+20\,*}}$   & 46\,$^{*}$      & -                 \\
        \cmidrule{2-9}
                         & \textbf{Total}    & 501            & 160                  & 208\,$^{\smash{+94\,*}}$      & 208\,$^{\smash{+101\,*}}$    & \textbf{217}\,$^{\smash{+103\,*}}$ & 324\,$^{*}$     & -                 \\
        \midrule
        \multirow{3}{*}{\shortstack[l]{Defects4J                                                                                                                                                                               \\multi-hunk\\(vs. BIRCH)}}
                         & SFMH              & 241            & 29                   & 54                            & 51                           & \textbf{58}                        & -               & 56                \\
                         & MFMH              & 126            & 15                   & 17                            & 18                           & 17                                 & -               & \textbf{24}       \\
        \cmidrule{2-9}
                         & \textbf{Total}    & 367            & 44                   & 71                            & 69                           & 75                                 & -               & \textbf{80}       \\
        \bottomrule
      \end{tabular}}
  \end{subtable}
\end{table}

\textbf{Using the same underlying LLM (DeepSeek-V3.2-Exp), \appname outperforms both state-of-the-art baselines, with its best configuration beating RepairAgent by +56.6\% and BIRCH-feedback by +47.1\%.}
On all 854 Defects4J bugs (Table~\ref{tab:rq2-ablation-deepseek}), the best \appname configuration (\fldiff) repairs 545 bugs versus 348 for RepairAgent, a +56.6\% improvement. Even \appname-\nonhistory (522), which uses no historical context, significantly outperforms RepairAgent by +50.0\% (McNemar's exact test, $p<0.001$), as does every other configuration. This margin over RepairAgent therefore cannot be attributed to history. It reflects framework differences between the two agents: \appname lets the model act through free-form \texttt{bash} commands, whereas RepairAgent drives a fixed set of repair tools with a finite-state machine. The margin is also not an artifact of a weak baseline: RepairAgent's own paper reports 186 plausible Defects4J repairs with GPT-3.5-0125~\citep{bouzenia2024repairagent}, while our replication on DeepSeek-V3.2-Exp, the LLM \appname also uses, increases this to 348.

On the 371 multi-hunk bugs evaluated by BIRCH-feedback (Table~\ref{tab:rq2-ablation-deepseek}), the best \appname configuration (\fnpair) repairs 175 bugs versus 119 for BIRCH-feedback, a +47.1\% improvement. By category, \appname-\fnpair repairs 133 SFMH and 42 MFMH bugs, compared with 85 and 34 for BIRCH-feedback. Here too, \appname-\nonhistory alone (172) already outperforms BIRCH-feedback (McNemar's exact test, $p<0.001$) and is the best configuration on MFMH (49 bugs), so this gap likewise cannot be read as evidence for history.

\textbf{On the two smaller models (Qwen3-Coder-Next and Devstral-Small-2), \appname's best configuration matches or beats RepairAgent.} We replicate the same-LLM baseline comparison (RepairAgent and BIRCH-feedback) on both smaller models (Tables~\ref{tab:rq2-ablation-qwen} and~\ref{tab:rq2-ablation-devstral}), testing each with McNemar's exact test. On Qwen3-Coder-Next, RepairAgent matches \appname's best configuration across all 854 Defects4J bugs (207 versus 212, no significant difference, $p=0.80$) but significantly outperforms \nonhistory (129, $p<0.001$). On Devstral-Small-2, \appname's best configuration significantly outperforms RepairAgent (330 versus 252, $p<0.001$), while RepairAgent still beats \nonhistory (218, $p=0.03$).

On the multi-hunk bugs, BIRCH-feedback matches \appname's best configuration on both Qwen (60 versus 54, $p=0.58$) and Devstral (80 versus 75, $p=0.70$). A likely reason the baselines stay competitive on these two models, having significantly underperformed \appname on DeepSeek, is the agent interface. RepairAgent and BIRCH-feedback issue structured function calls through the models' native tool-calling APIs, while \appname has the model generate free-form \texttt{bash} commands that our harness parses. This interface difference is particularly relevant to Qwen3-Coder-Next, which is optimized for tool use. More importantly, the difference arises across frameworks rather than within \appname: under the same fixed interface, the best history configuration consistently outperforms \nonhistory across all three models.

\textbf{The benefit of an individual history heuristic grows as the base model weakens: modest and not significant on DeepSeek, but large and significant on the two smaller models.}
We run the four-configuration comparison on all three base models under perfect FL, on all 854 Defects4J and 501 BugsInPy bugs, using the identical protocol, prompts, and Docker harness across models (Tables~\ref{tab:rq2-ablation-deepseek}, \ref{tab:rq2-ablation-qwen}, and~\ref{tab:rq2-ablation-devstral}). On the strong DeepSeek model the best history configuration improves Plausible@1 over \nonhistory by only 4.4\% on Defects4J (545 versus 522) and 1.1\% on BugsInPy (377 versus 373), and no individual history configuration significantly beats \nonhistory: even the best of the three is not significant under McNemar's exact test ($p = 0.12$ on Defects4J and $p = 0.79$ on BugsInPy), unlike \appname's significant margins over RepairAgent and BIRCH-feedback. On the two smaller models, the same comparison is far larger and significant: the best history configuration beats \nonhistory by $+51.4\%$ (Defects4J) and $+35.6\%$ (BugsInPy) on Devstral, and $+64.3\%$ and $+31.9\%$ on Qwen, each significant on both benchmarks (all $p<0.01$). On Defects4J, the gain grows steadily as the model weakens, from $+4.4\%$ (DeepSeek) to $+51.4\%$ (Devstral) to $+64.3\%$ (Qwen): a stronger model fixes more bugs from the buggy snapshot alone, while a weaker model relies on history to fill gaps in its understanding of the codebase. Because both smaller models are served locally, this benefit is reproducible at no API cost. On DeepSeek, where the aggregate lift is modest, the value of history instead lies in the complementarity of the three heuristics, which we examine next.

\begin{figure}[t]
  \centering
  \includegraphics[width=\textwidth]{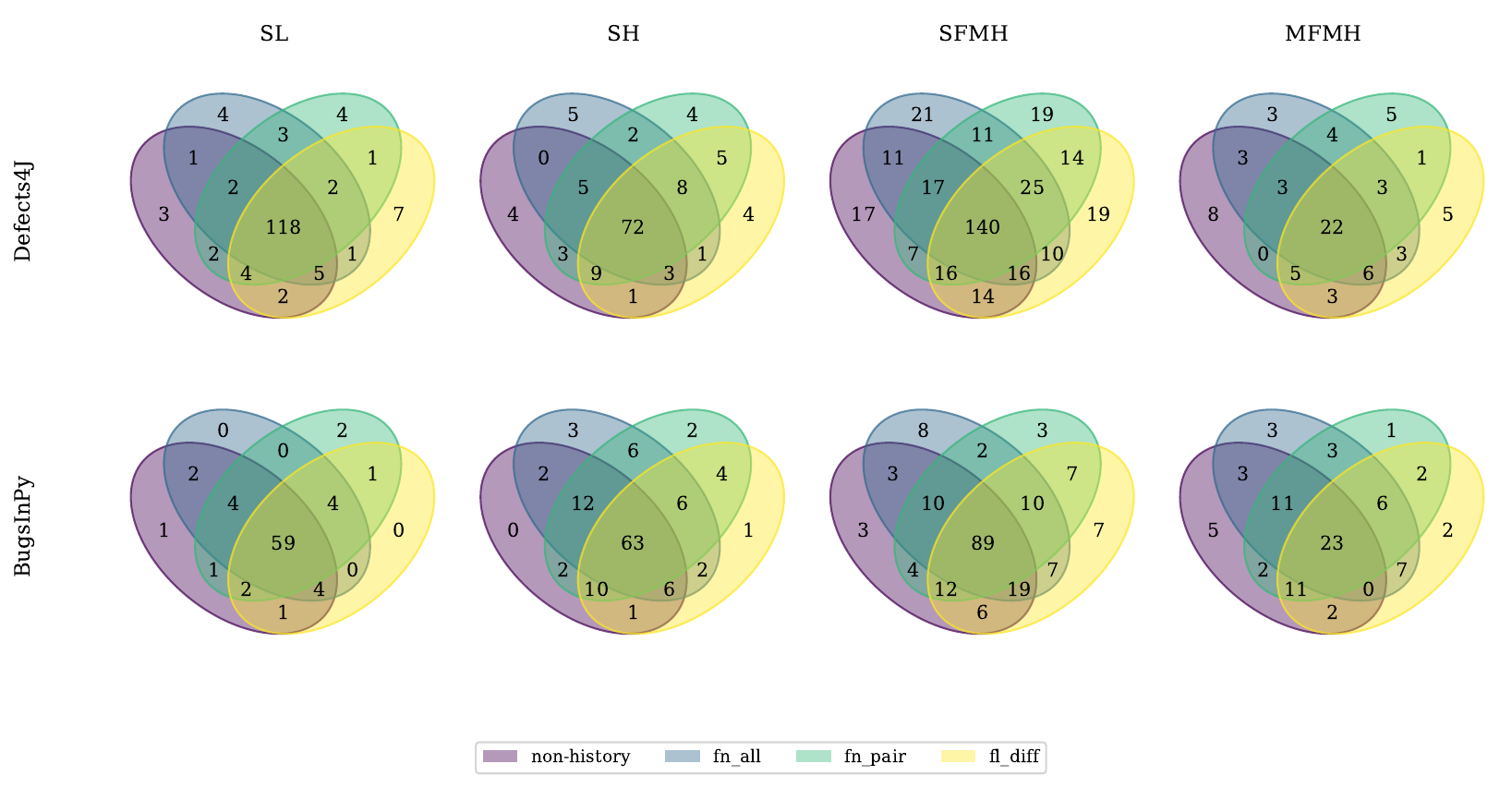}
  \caption{Venn diagrams comparing bug repair overlap across the four context configurations on DeepSeek-V3.2-Exp, on Defects4J (top) and BugsInPy (bottom). Each number shows the count of bugs uniquely fixed by that intersection of configurations.}
  \label{fig:rq1_venn_combined}
\end{figure}

\textbf{The three history heuristics are complementary, solving distinct sets of bugs.} Figure~\ref{fig:rq1_venn_combined} shows that in each category the four configurations agree on a large common set of bugs, yet each history heuristic also repairs a nontrivial number that the others miss. On DeepSeek, each history heuristic uniquely repairs 106 to 111 Defects4J and 59 to 67 BugsInPy bugs that \nonhistory does not, while \nonhistory in turn repairs a comparable number that history misses. On Defects4J SFMH, \fnall, \fnpair, and \fldiff each add 21, 19, and 19 unique fixes beyond the 140-bug common set. On BugsInPy SFMH, the four configurations share 89 bugs, yet each history heuristic repairs 22 to 31 that \nonhistory misses, lifting the hybrid of the three heuristics to 187 versus 146 for \nonhistory. Even on MFMH, where \nonhistory has the highest total, the history configurations collectively contribute 24 unique fixes versus 8 from \nonhistory. Because all configurations share the same LLM, prompt scaffold, and temperature~0, these unique repairs are attributable to the injected historical context rather than sampling variation: history does not simply add fixes on top of \nonhistory, but repairs a comparably large and distinct set of bugs. This confirms that \fnall (function-name co-evolution), \fnpair (before/after semantics), and \fldiff (fine-grained edits) capture complementary signals.

\textbf{A hybrid approach comprising the union of the three history heuristics repairs statistically significantly more bugs than \nonhistory.} The \textbf{Hybrid} column of Table~\ref{tab:rq2-ablation-deepseek} marks with $^{*}$ each cell where the hybrid approach, the union of the three heuristics that counts a bug as repaired if at least one heuristic fixes it, significantly repairs bugs \nonhistory cannot (McNemar's exact test, paired per bug). The hybrid beats \nonhistory in seven of eight category cells and overall on both benchmarks, because it collects the different bugs each heuristic uniquely rescues. Pooled across all bugs, the hybrid approach repairs 684 of 854 Defects4J bugs (80.1\%) versus 522 (61.1\%) for \nonhistory and 463 of 501 BugsInPy bugs (92.4\%) versus 373 (74.5\%), beating \nonhistory with $p<0.001$ on both benchmarks. This is the statistical counterpart of the overlaps in Figure~\ref{fig:rq1_venn_combined}: because each heuristic repairs a different subset of the bugs \nonhistory misses, combining all three significantly increases the number of bugs repaired. In practice, a practitioner could realize this combined set either by \emph{brute force}, running all three history heuristics and keeping any test-passing patch, or by a cheaper \emph{cascade} that runs them in a fixed order and stops at the first test-passing patch, at most three attempts per bug. Both variants reach the same 684 and 463 repairs. We analyze the cost of the brute-force and cascade strategies for the history hybrid in RQ3 (Section~\ref{subsec:RQ3}).

Because Plausible@1 counts test-passing patches, which are not guaranteed to be correct fixes, we manually verify a stratified sample and find that the history advantage holds under correctness as well, not only test-passing (Section~\ref{sec:discussion-and-future-work}, ``Plausible versus Correct Patches'').

\begin{rqbox}{Summary for RQ1}
  \begin{enumerate}
    \item On the same LLM, \appname beats RepairAgent (+56.6\%) and BIRCH-feedback (+47.1\%), but so does \nonhistory, so this gap measures the agent framework, not history. History's effect is isolated by the history-vs-\nonhistory comparison.
    \item The best history configuration beats \nonhistory on all three base models under perfect FL, and more so as the model weakens: $+4.4\%$/$+1.1\%$ on DeepSeek (not significant), $+51.4\%$/$+35.6\%$ on Devstral, and $+64.3\%$/$+31.9\%$ on Qwen ($p<0.01$, Defects4J/BugsInPy).
    \item On DeepSeek, no individual heuristic is significant, yet each one uniquely repairs 106 to 111 (Defects4J) and 59 to 67 (BugsInPy) bugs that \nonhistory misses, and the hybrid approach significantly beats \nonhistory ($p<0.001$).\end{enumerate}
\end{rqbox}

\subsection{RQ2: \rqtwo}\label{subsec:RQ2}
\subsubsection{Motivation}\label{subsubsec:RQ2_Motivation}
All experiments so far assume perfect fault localization, which is unrealistic in practice~\citep{yang2024sweagent,fan2023automated}: real FL tools produce imprecise locations. Hence, when blame is computed on incorrect lines, irrelevant commits may be retrieved by our approach, potentially negating the history benefit. While perfect FL remains the right setting for isolating history's effect in RQ1, because it removes localization as a confound, this RQ complements that controlled result with a realism check: it re-runs all configurations under spectrum-based fault localization (SBFL), an automated technique that ranks suspicious code from test coverage, to test whether the history benefit observed under perfect localization survives the imperfect locations a practitioner would actually obtain. Prior history-aware APR assumes perfect FL~\citep{shi2025hafix}, while the closest agentic system, RepairAgent, reports realistic FL only for a 100-bug Defects4J subset and only for overall repair~\citep{bouzenia2024repairagent}. Our empirical evaluation pairs perfect and realistic FL on the same bugs, all 827 Defects4J and 361 BugsInPy bugs for which SBFL produces a ranking, across two languages and three base models.

\subsubsection{Approach}\label{subsubsec:RQ2_Approach}

\begin{table}[t]
  \centering
  \caption{Accuracy for the realistic spectrum-based fault localization (SBFL) used in RQ2 (Ochiai formula, line granularity). \#Bugs is the number of SBFL-runnable bugs. Top-$N$ is the number (percentage) of those bugs for which at least one developer-patched line is ranked within the top~$N$ suspicious lines.}
  \label{tab:fl_accuracy}
  \small
  \begin{tabular}{llccccc}
    \toprule
    Benchmark & FL tool       & \#Bugs & Top-1    & Top-3      & Top-5      & Top-10     \\
    \midrule
    Defects4J & GZoltar 1.7.3 & 827    & 40 (5\%) & 137 (17\%) & 183 (22\%) & 245 (30\%) \\
    BugsInPy  & FauxPy 0.7.0  & 361    & 13 (4\%) & 54 (15\%)  & 71 (20\%)  & 107 (30\%) \\
    \bottomrule
  \end{tabular}
\end{table}

We run one widely used SBFL tool per language, which executes the test suite and ranks the program's source lines by how likely each is to be faulty, and feed the top of that ranked list into the same repair pipeline used in RQ1. For Defects4J we use GZoltar~1.7.3~\citep{DBLP:conf/taicpart/RiboiraA10} and for BugsInPy we use FauxPy~0.7.0~\citep{DBLP:journals/corr/abs-2404-18596}. Both score each line with the Ochiai suspiciousness formula, which rewards lines that are executed by many failing tests and few passing ones, and both compute this score at line (single source-line) granularity rather than at the coarser method or file level, so the agent receives specific candidate lines. We run the agent once per bug. All ten ranked lines replace RQ1's developer-patched lines as the agent's fault locations (the same set across all four configurations). As in RQ1, the LLM judge runs \texttt{git blame} on one of the ten, and the single commit it returns supplies the history context for the three history configurations. The locations modified by the developers' golden patch are used only post hoc to measure success and FL accuracy, never in the prompt or in blame. Table~\ref{tab:fl_accuracy} illustrates the realism of this localization: a developer-modified line appears in the top-10 for only ${\sim}30\%$ of bugs on both benchmarks, confirming that the localization is substantially noisier than the perfect FL assumed in prior work.

We run all four configurations on every bug for which the SBFL tool produces a ranking, i.e., 827 of 854 Defects4J bugs and 361 of 501 BugsInPy bugs, and compare against the perfect-FL results from RQ1 on the same bugs. We run this SBFL evaluation on all three base models and, below, contrast DeepSeek-V3.2-Exp with the two smaller models directly. The four structural bug categories (SL, SH, SFMH, and MFMH, defined in Section~\ref{subsubsec:bug-category-terminology} in terms of the number and location of hunks in the developer patch) are used only to group results after the fact, and are never provided to the agent under SBFL. We report the Plausible@1 success rate here. Cost and agent effort under SBFL are analyzed alongside the perfect-FL setting in RQ3 (Section~\ref{subsec:RQ3}). Table~\ref{tab:rq3-sbfl-crossmodel} reports the full per-category SBFL results for all three models.



\begin{table}[t]
  \centering
  \caption{Repair effectiveness under realistic spectrum-based fault localization (SBFL) across the three base models, paired on the FL-runnable bugs per dataset (827 Defects4J, 361 BugsInPy). Each cell is the Plausible@1 success rate (\%) under SBFL. The subscript is the change in percentage points relative to perfect FL on the same bugs (Table~\ref{tab:rq2-ablation} provides perfect-FL results on the full benchmarks). \textbf{Bold} marks the best Overall rate per dataset among the four configurations. Category sizes (SL/SH/SFMH/MFMH) are 160/131/413/123 for Defects4J and 51/90/150/70 for BugsInPy.}
  \label{tab:rq3-sbfl-crossmodel}
  \small
  \setlength{\tabcolsep}{5pt}

  \begin{subtable}[t]{\linewidth}
    \centering
    \caption{DeepSeek-V3.2-Exp}
    \label{tab:rq3_repair}
    \begin{tabular}{ll ccccc}
      \toprule
      Dataset & Config      & Overall                   & SL           & SH           & SFMH         & MFMH         \\
      \midrule
      \multirow{4}{*}{Defects4J}
              & \nonhistory & 56.3\%$_{-4.2}$           & 73\%$_{-8}$  & 57\%$_{-15}$ & 57\%$_{+1}$  & 32\%$_{-7}$  \\
              & \fnall      & 58.2\%$_{-3.6}$           & 73\%$_{-8}$  & 65\%$_{-6}$  & 56\%$_{-3}$  & 40\%$_{+4}$  \\
              & \fnpair     & \textbf{58.4\%$_{-3.6}$}  & 76\%$_{-4}$  & 68\%$_{-12}$ & 56\%$_{-2}$  & 34\%$_{+2}$  \\
              & \fldiff     & 57.7\%$_{-5.3}$           & 71\%$_{-12}$ & 57\%$_{-19}$ & 61\%$_{+1}$  & 31\%$_{-5}$  \\
      \midrule
      \multirow{4}{*}{BugsInPy}
              & \nonhistory & 57.9\%$_{-16.6}$          & 71\%$_{-20}$ & 62\%$_{-16}$ & 55\%$_{-15}$ & 49\%$_{-20}$ \\
              & \fnall      & 58.4\%$_{-15.0}$          & 71\%$_{-18}$ & 61\%$_{-17}$ & 58\%$_{-15}$ & 47\%$_{-11}$ \\
              & \fnpair     & 57.9\%$_{-16.1}$          & 71\%$_{-16}$ & 62\%$_{-21}$ & 52\%$_{-16}$ & 56\%$_{-10}$ \\
              & \fldiff     & \textbf{63.7\%$_{-11.1}$} & 75\%$_{-16}$ & 69\%$_{-4}$  & 63\%$_{-17}$ & 51\%$_{-4}$  \\
      \bottomrule
    \end{tabular}
  \end{subtable}

  \vspace{0.6em}

  \begin{subtable}[t]{\linewidth}
    \centering
    \caption{Qwen3-Coder-Next}
    \label{tab:app-qwen-sbfl}
    \begin{tabular}{ll ccccc}
      \toprule
      Dataset & Config      & Overall                  & SL           & SH           & SFMH         & MFMH        \\
      \midrule
      \multirow{4}{*}{Defects4J}
              & \nonhistory & 10.5\%$_{-4.5}$          & 21\%$_{-5}$  & 10\%$_{+0}$  & 9\%$_{-6}$   & 4\%$_{-5}$  \\
              & \fnall      & 13.8\%$_{-6.0}$          & 24\%$_{-9}$  & 15\%$_{-2}$  & 12\%$_{-8}$  & 6\%$_{-1}$  \\
              & \fnpair     & 13.3\%$_{-11.2}$         & 24\%$_{-15}$ & 11\%$_{-18}$ & 12\%$_{-10}$ & 5\%$_{-4}$  \\
              & \fldiff     & \textbf{17.5\%$_{-4.6}$} & 29\%$_{-8}$  & 19\%$_{-3}$  & 17\%$_{-4}$  & 2\%$_{-5}$  \\
      \midrule
      \multirow{4}{*}{BugsInPy}
              & \nonhistory & \textbf{15.0\%$_{-0.6}$} & 31\%$_{-4}$  & 19\%$_{-3}$  & 12\%$_{+1}$  & 4\%$_{+3}$  \\
              & \fnall      & 10.0\%$_{-8.3}$          & 24\%$_{-18}$ & 12\%$_{-11}$ & 9\%$_{-5}$   & 0\%$_{-4}$  \\
              & \fnpair     & 14.1\%$_{-6.9}$          & 22\%$_{-24}$ & 17\%$_{-2}$  & 11\%$_{-9}$  & 11\%$_{+4}$ \\
              & \fldiff     & 13.0\%$_{-10.0}$         & 22\%$_{-25}$ & 14\%$_{-16}$ & 15\%$_{-5}$  & 1\%$_{-3}$  \\
      \bottomrule
    \end{tabular}
  \end{subtable}

  \vspace{0.6em}

  \begin{subtable}[t]{\linewidth}
    \centering
    \caption{Devstral-Small-2}
    \label{tab:app-devstral-sbfl}
    \begin{tabular}{ll ccccc}
      \toprule
      Dataset & Config      & Overall                  & SL           & SH           & SFMH         & MFMH         \\
      \midrule
      \multirow{4}{*}{Defects4J}
              & \nonhistory & 34.0\%$_{+8.9}$          & 52\%$_{-2}$  & 37\%$_{+13}$ & 30\%$_{+13}$ & 20\%$_{+7}$  \\
              & \fnall      & 34.2\%$_{+1.5}$          & 56\%$_{+3}$  & 38\%$_{-5}$  & 29\%$_{+2}$  & 20\%$_{+5}$  \\
              & \fnpair     & 35.1\%$_{+3.0}$          & 52\%$_{-4}$  & 37\%$_{-2}$  & 33\%$_{+8}$  & 18\%$_{+2}$  \\
              & \fldiff     & \textbf{37.0\%$_{-1.2}$} & 54\%$_{-13}$ & 46\%$_{-5}$  & 32\%$_{+1}$  & 24\%$_{+11}$ \\
      \midrule
      \multirow{4}{*}{BugsInPy}
              & \nonhistory & 41.6\%$_{+7.8}$          & 51\%$_{+6}$  & 50\%$_{+10}$ & 38\%$_{+5}$  & 31\%$_{+13}$ \\
              & \fnall      & \textbf{44.3\%$_{+0.6}$} & 55\%$_{-6}$  & 54\%$_{+1}$  & 40\%$_{-2}$  & 33\%$_{+10}$ \\
              & \fnpair     & 41.3\%$_{-0.3}$          & 51\%$_{-4}$  & 51\%$_{+2}$  & 40\%$_{+3}$  & 24\%$_{-7}$  \\
              & \fldiff     & 42.7\%$_{-1.9}$          & 53\%$_{-8}$  & 51\%$_{-6}$  & 37\%$_{+0}$  & 36\%$_{+3}$  \\
      \bottomrule
    \end{tabular}
  \end{subtable}
\end{table}

\subsubsection{Results}\label{subsubsec:RQ2_Results}

\textbf{Under realistic SBFL, the improvement that individual history heuristics showed over \nonhistory (especially for weaker base models) under perfect FL disappears: the best individual configuration significantly beats \nonhistory in only one of six model-benchmark cells, i.e., Qwen on Defects4J.} On DeepSeek, neither margin is significant, similar to under perfect FL: \fnpair reaches $58.4\%$ against $56.3\%$ for \nonhistory on Defects4J ($+2.1$ percentage points) and \fldiff reaches $63.7\%$ against $57.9\%$ on BugsInPy ($+5.8$), and neither passes McNemar's test ($p = 0.33$ and $p = 0.078$, Table~\ref{tab:rq3_repair}). The larger perfect-FL gains on the two smaller models ($+51.4\%$/$+35.6\%$ on Devstral, $+64.3\%$/$+31.9\%$ on Qwen, all $p<0.01$) mostly vanish under noisy localization (Tables~\ref{tab:app-qwen-sbfl} and~\ref{tab:app-devstral-sbfl}). On Qwen, the best history success rate falls from $24.5\%$ to $17.5\%$ on Defects4J, where it still significantly beats \nonhistory ($10.5\%$), but on BugsInPy it falls from $23.0\%$ to $14.1\%$ while \nonhistory barely moves ($15.5\%$ to $15.0\%$), so on that benchmark \nonhistory is the strongest individual configuration and no history heuristic helps. On Devstral, the best history configuration keeps a higher but not significant success rate than \nonhistory on both benchmarks, about three percentage points (Defects4J \fldiff $37.0\%$ versus $34.0\%$, BugsInPy \fnall $44.3\%$ versus $41.6\%$, with $p = 0.084$ and $p = 0.395$), while \nonhistory itself does better under SBFL than under perfect FL ($+8.9$ and $+7.8$ percentage points). This is possibly because under SBFL \nonhistory is given the ten ranked suspicious lines as fault locations, more candidate sites than the exact patched lines it receives under perfect FL.

\begin{table}[p]
  \centering
  \caption{History versus non-history under realistic spectrum-based fault localization (SBFL) across the three base models, paired on the FL-runnable bugs per dataset (827 Defects4J, 361 BugsInPy, the same bugs for every model). Each cell is \#Pass, i.e., the number of bugs plausibly repaired (the Plausible@1 rate is \#Pass$/$Total). 
  Superscripts count the bugs a history configuration repairs that \nonhistory does not.
  \textbf{Hybrid} $=$ number of bugs fixed by at least one of the three history heuristics (\fnall, \fnpair, \fldiff). 
  \textbf{Bold} $=$ best individual configuration per category, including \nonhistory but excluding the Hybrid. 
  A $*$ marks a history heuristic or Hybrid that repairs significantly more bugs than \nonhistory (McNemar's exact test versus \nonhistory, Bonferroni $\alpha = 0.05/4 = 0.0125$ across the three heuristics and their union). Category sizes (SL/SH/SFMH/MFMH) are 160/131/413/123 for Defects4J and 51/90/150/70 for BugsInPy.}
  \label{tab:rq4-ablation}
  \scriptsize
  \setlength{\tabcolsep}{3pt}
  \renewcommand{\arraystretch}{0.70}

  \begin{subtable}[t]{\linewidth}
    \centering
    \caption{DeepSeek-V3.2-Exp}
    \label{tab:rq4-ablation-deepseek}
    \resizebox{0.78\textwidth}{!}{%
      \begin{tabular}{l l r r r r r r}
        \toprule
        \textbf{Dataset}           & \textbf{Category} & \textbf{Total} & \textbf{\nonhistory} & \textbf{\fnall}       & \textbf{\fnpair}        & \textbf{\fldiff}       & \textbf{Hybrid} \\
        \midrule
        \multirow{5}{*}{Defects4J} & SL                & 160            & 117                  & 117\,$^{+18}$         & \textbf{122}\,$^{+17}$  & 114\,$^{+20}$          & 144\,$^{*}$     \\
                                   & SH                & 131            & 75                   & 85\,$^{+28}$          & \textbf{89}\,$^{+29}$   & 75\,$^{+20}$           & 115\,$^{*}$     \\
                                   & SFMH              & 413            & 235                  & 230\,$^{+68}$         & 230\,$^{+78}$           & \textbf{250}\,$^{+81}$ & 344\,$^{*}$     \\
                                   & MFMH              & 123            & 39                   & \textbf{49}\,$^{+24}$ & 42\,$^{+21}$            & 38\,$^{+21}$           & 76\,$^{*}$      \\
        \cmidrule(lr){2-8}
                                   & \textbf{Total}    & 827            & 466                  & 481\,$^{+138}$        & \textbf{483}\,$^{+145}$ & 477\,$^{+142}$         & 679\,$^{*}$     \\
        \midrule
        \multirow{5}{*}{BugsInPy}  & SL                & 51             & 36                   & 36\,$^{+10}$          & 36\,$^{+8}$             & \textbf{38}\,$^{+8}$   & 44              \\
                                   & SH                & 90             & 56                   & 55\,$^{+13}$          & 56\,$^{+17}$            & \textbf{62}\,$^{+21}$  & 78\,$^{*}$      \\
                                   & SFMH              & 150            & 83                   & 87\,$^{+26}$          & 78\,$^{+21}$            & \textbf{94}\,$^{+33}$  & 123\,$^{*}$     \\
                                   & MFMH              & 70             & 34                   & 33\,$^{+14}$          & \textbf{39}\,$^{+18}$   & 36\,$^{+13}$           & 54\,$^{*}$      \\
        \cmidrule(lr){2-8}
                                   & \textbf{Total}    & 361            & 209                  & 211\,$^{+63}$         & 209\,$^{+64}$           & \textbf{230}\,$^{+75}$ & 299\,$^{*}$     \\
        \bottomrule
      \end{tabular}}
  \end{subtable}

  \vspace{0.1em}

  \begin{subtable}[t]{\linewidth}
    \centering
    \caption{Qwen3-Coder-Next}
    \label{tab:rq4-ablation-qwen}
    \resizebox{0.78\textwidth}{!}{%
      \begin{tabular}{l l r r r r r r}
        \toprule
        \textbf{Dataset}           & \textbf{Category} & \textbf{Total} & \textbf{\nonhistory} & \textbf{\fnall}     & \textbf{\fnpair}    & \textbf{\fldiff}          & \textbf{Hybrid} \\
        \midrule
        \multirow{5}{*}{Defects4J} & SL                & 160            & 33                   & 38\,$^{+19}$        & 39\,$^{+15}$        & \textbf{47}\,$^{+24}$     & 63\,$^{*}$      \\
                                   & SH                & 131            & 13                   & 19\,$^{+12}$        & 14\,$^{+7}$         & \textbf{25}\,$^{+18}$     & 37\,$^{*}$      \\
                                   & SFMH              & 413            & 36                   & 50\,$^{+29}$        & 51\,$^{+31}$        & \textbf{70}\,$^{+46\,*}$  & 101\,$^{*}$     \\
                                   & MFMH              & 123            & 5                    & \textbf{7}\,$^{+3}$ & 6\,$^{+4}$          & 3\,$^{+2}$                & 11              \\
        \cmidrule(lr){2-8}
                                   & \textbf{Total}    & 827            & 87                   & 114\,$^{+63\,*}$    & 110\,$^{+57}$       & \textbf{145}\,$^{+90\,*}$ & 212\,$^{*}$     \\
        \midrule
        \multirow{5}{*}{BugsInPy}  & SL                & 51             & \textbf{16}          & 12\,$^{+4}$         & 11\,$^{+2}$         & 11\,$^{+2}$               & 18              \\
                                   & SH                & 90             & \textbf{17}          & 11\,$^{+5}$         & 15\,$^{+6}$         & 13\,$^{+7}$               & 27              \\
                                   & SFMH              & 150            & 18                   & 13\,$^{+8}$         & 17\,$^{+10}$        & \textbf{22}\,$^{+11}$     & 32\,$^{*}$      \\
                                   & MFMH              & 70             & 3                    & 0\,$^{+0}$          & \textbf{8}\,$^{+6}$ & 1\,$^{+0}$                & 8               \\
        \cmidrule(lr){2-8}
                                   & \textbf{Total}    & 361            & \textbf{54}          & 36\,$^{+17}$        & 51\,$^{+24}$        & 47\,$^{+20}$              & 85\,$^{*}$      \\
        \bottomrule
      \end{tabular}}
  \end{subtable}

  \vspace{0.1em}

  \begin{subtable}[t]{\linewidth}
    \centering
    \caption{Devstral-Small-2}
    \label{tab:rq4-ablation-devstral}
    \resizebox{0.78\textwidth}{!}{%
      \begin{tabular}{l l r r r r r r}
        \toprule
        \textbf{Dataset}           & \textbf{Category} & \textbf{Total} & \textbf{\nonhistory} & \textbf{\fnall}        & \textbf{\fnpair}       & \textbf{\fldiff}        & \textbf{Hybrid} \\
        \midrule
        \multirow{5}{*}{Defects4J} & SL                & 160            & 84                   & \textbf{89}\,$^{+21}$  & 84\,$^{+16}$           & 86\,$^{+20}$            & 115\,$^{*}$     \\
                                   & SH                & 131            & 49                   & 50\,$^{+17}$           & 49\,$^{+16}$           & \textbf{60}\,$^{+24}$   & 86\,$^{*}$      \\
                                   & SFMH              & 413            & 124                  & 120\,$^{+43}$          & \textbf{135}\,$^{+59}$ & 131\,$^{+54}$           & 210\,$^{*}$     \\
                                   & MFMH              & 123            & 24                   & 24\,$^{+9}$            & 22\,$^{+7}$            & \textbf{29}\,$^{+11}$   & 38\,$^{*}$      \\
        \cmidrule(lr){2-8}
                                   & \textbf{Total}    & 827            & 281                  & 283\,$^{+90}$          & 290\,$^{+98}$          & \textbf{306}\,$^{+109}$ & 449\,$^{*}$     \\
        \midrule
        \multirow{5}{*}{BugsInPy}  & SL                & 51             & 26                   & \textbf{28}\,$^{+9}$   & 26\,$^{+7}$            & 27\,$^{+10}$            & 37              \\
                                   & SH                & 90             & 45                   & \textbf{49}\,$^{+16}$  & 46\,$^{+10}$           & 46\,$^{+12}$            & 63\,$^{*}$      \\
                                   & SFMH              & 150            & 57                   & \textbf{60}\,$^{+26}$  & \textbf{60}\,$^{+28}$  & 56\,$^{+23}$            & 89\,$^{*}$      \\
                                   & MFMH              & 70             & 22                   & 23\,$^{+10}$           & 17\,$^{+5}$            & \textbf{25}\,$^{+9}$    & 36\,$^{*}$      \\
        \cmidrule(lr){2-8}
                                   & \textbf{Total}    & 361            & 150                  & \textbf{160}\,$^{+61}$ & 149\,$^{+50}$          & 154\,$^{+54}$           & 225\,$^{*}$     \\
        \bottomrule
      \end{tabular}}
  \end{subtable}
\end{table}

\textbf{The three history heuristics remain complementary under realistic SBFL on all three base models, and on DeepSeek the hybrid approach beats \nonhistory by a wider margin than under perfect FL.} On DeepSeek (Table~\ref{tab:rq4-ablation-deepseek}) each history heuristic still uniquely repairs bugs \nonhistory misses under SBFL, 138 to 145 on Defects4J and 63 to 75 on BugsInPy, so the hybrid approach comprising their union fixes 679 Defects4J and 299 BugsInPy bugs against 466 and 209 for \nonhistory. This margin is wider under SBFL than under perfect FL ($+213$ versus $+159$ bugs on Defects4J, $+90$ versus $+62$ on BugsInPy): as localization degrades, \nonhistory loses more repairs than the hybrid approach does. Each heuristic's unique fixes are offset by a comparable number of \nonhistory-only fixes, yet the hybrid significantly beats \nonhistory in seven of eight category cells and overall (McNemar's exact test, $p<0.001$, marked $^{*}$ in the table) on both benchmarks. The hybrid approach stays ahead on the smaller models too: it significantly beats \nonhistory in every overall cell for Qwen and Devstral (Tables~\ref{tab:rq4-ablation-qwen} and~\ref{tab:rq4-ablation-devstral}, $p<0.001$), and on Qwen BugsInPy it repairs 85 bugs against 54 for \nonhistory even though \nonhistory is the strongest individual configuration there.

The individual heuristics stop helping the weaker models once localization is imperfect. This likely happens because a wrong SBFL ranking sends \texttt{git blame} to a non-faulty line, which returns an unrelated commit, hence the injected history points the model at the wrong code. The large DeepSeek model can set such misleading history aside, but the weaker models tend to follow it. Hence, under realistic localization, it is the hybrid approach, not any individual heuristic, that generalizes across models: even on the weaker models, and even where \nonhistory is the strongest individual configuration, the three heuristics repair enough distinct bugs that the hybrid still repairs more than \nonhistory.

\begin{rqbox}{Summary for RQ2}
  \begin{enumerate}
    \item Realistic SBFL is far noisier than the perfect FL assumed in prior work: on both benchmarks, only about 30\% of bugs have a developer-modified line in the top-10 ranked suspicious lines.
    \item Under realistic SBFL, the single-heuristic gain that grew as the base model weakened under perfect FL disappears: the best individual configuration significantly beats \nonhistory in only one of six model-benchmark cells, i.e., Qwen on Defects4J. On DeepSeek, improvements of the history heuristics are not significant under SBFL ($p = 0.33$ and $p = 0.078$), as they were not under perfect FL.
    \item The three heuristics stay complementary on all three models: the hybrid significantly beats \nonhistory in every overall cell ($p<0.001$), while on DeepSeek the hybrid's margin over \nonhistory is wider under SBFL than under perfect FL ($+213$ versus $+159$ bugs on Defects4J, $+90$ versus $+62$ on BugsInPy).
  \end{enumerate}
\end{rqbox}

\subsection{RQ3: \rqthree}\label{subsec:RQ3}

\subsubsection{Motivation}\label{subsubsec:RQ3_Motivation}
While RQ1 shows that historical context improves repair effectiveness, cost is a primary concern for agent systems, as iterative tool use can accumulate substantial token and time budgets, even on simple bugs~\citep{bouzenia2024repairagent,xia2024automated}. Recent evaluations also highlight the risk of expensive failures where unresolved attempts consume several times more tokens than successful ones~\citep{DBLP:journals/corr/abs-2509-09853}. Since \appname integrates additional contextual information into each prompt, it is important to verify that the performance gains do not incur significant cost overhead, both under perfect fault localization and under the realistic spectrum-based fault localization (SBFL) setting of RQ2. This RQ asks whether adding historical context inflates agent cost or reasoning steps relative to \nonhistory.

\subsubsection{Approach}\label{subsubsec:RQ3_Approach}
We evaluate inference cost (USD) and agent reasoning steps across all four configurations on Defects4J (854 bugs) and BugsInPy (501 bugs). Inference cost is computed from token usage converted to USD using API pricing rates for DeepSeek-V3.2-Exp. We report cost for DeepSeek-V3.2-Exp only, because it is the only one of our three base models priced through a metered API. The other two run locally and have no comparable per-token cost. A run terminates upon a successful repair or by hitting one of the predefined limits (50 steps, \$1~USD cost, or 1-hour timeout), as defined in Section~\ref{subsec:Implementation}. To ensure fair comparison, we report cost and step metrics on the subset of bugs fixed by all four configurations, and assess statistical significance using Friedman tests followed by pairwise Wilcoxon signed-rank tests with Bonferroni correction. We report both the perfect-FL setting and, along the FL axis, a paired perfect-versus-SBFL comparison on the bugs each configuration solves in both regimes.

Finally, we price the \emph{hybrid} approach, i.e., the union of the three history heuristics, on the same DeepSeek runs. It has two variants. The \emph{brute-force} variant runs all three heuristics on every bug and keeps any test-passing patch, always paying for three runs. The \emph{cascade} variant runs them in the order they are defined in Section~\ref{subsubsec:historical-context}, \fnall then \fnpair then \fldiff, from the broadest context to the finest-grained edit, and stops at the first test-passing patch. Hence, this variant pays only for the heuristics it runs: if \fnall fails and \fnpair passes, the bug pays for \fnall and \fnpair only. The two variants repair exactly the same set of bugs, because a bug is repaired whenever any one of the three heuristics passes, so the order matters only for the cascade's cost, not for which bugs it fixes. We test each variant's per-bug cost against \nonhistory with a paired Wilcoxon signed-rank test, reusing the completed runs.

\subsubsection{Results}\label{subsubsec:RQ3_Results}

\begin{table}[t]
  \centering
  \caption{Median cost (USD) and the number of agent steps for successful repairs on DeepSeek-V3.2-Exp under perfect fault localization, with Friedman test $p$-values for cost on the matched set (N = bugs fixed by all 4 configs). Bold $p$-values indicate significance ($p < 0.05$), and $^{*}$ marks configs with significantly higher cost than \nonhistory (pairwise Wilcoxon post-hoc, Bonferroni $\alpha = 0.0167$).}
  \label{tab:rq2b-cost-crosslang}
  \setlength{\tabcolsep}{4pt}
  \small
  \resizebox{\textwidth}{!}{%
    \begin{tabular}{l l r r r r r r r r r r}
      \toprule
                                 &               &            & \multicolumn{4}{c}{\textbf{Median Cost (USD)}} & \multicolumn{4}{c}{\textbf{Median Steps}}                                                                                        \\
      \cmidrule(lr){4-7} \cmidrule(lr){8-11}
      \textbf{Dataset}           & \textbf{Cat.} & \textbf{N} & \nonhistory                                    & \fnall                                    & \fnpair & \fldiff     & \nonhistory & \fnall & \fnpair & \fldiff & \textbf{Fried. p} \\
      \midrule
      \multirow{4}{*}{Defects4J} & SL            & 118        & 0.005                                          & 0.008$^{*}$                               & 0.005   & 0.008$^{*}$ & 12          & 14     & 13      & 16      & \textbf{$<$0.001} \\
                                 & SH            & 72         & 0.016                                          & 0.019                                     & 0.015   & 0.014       & 23          & 25     & 24      & 22      & 0.184             \\
                                 & SFMH          & 140        & 0.016                                          & 0.020                                     & 0.016   & 0.021       & 26          & 28     & 26      & 30      & 0.122             \\
                                 & MFMH          & 22         & 0.029                                          & 0.026                                     & 0.023   & 0.023       & 34          & 29     & 32      & 30      & 0.142             \\
      \midrule
      \multirow{4}{*}{BugsInPy}  & SL            & 59         & 0.027                                          & 0.029                                     & 0.020   & 0.033       & 19          & 19     & 18      & 18      & 0.214             \\
                                 & SH            & 63         & 0.030                                          & 0.031                                     & 0.034   & 0.036       & 23          & 18     & 23      & 21      & 0.252             \\
                                 & SFMH          & 89         & 0.053                                          & 0.037                                     & 0.055   & 0.055       & 28          & 24     & 24      & 23      & 0.062             \\
                                 & MFMH          & 23         & 0.062                                          & 0.064                                     & 0.066   & 0.060       & 29          & 30     & 31      & 30      & 0.304             \\
      \bottomrule
    \end{tabular}}
\end{table}

\textbf{Successful repairs finish well under the 50-step limit.}
Table~\ref{tab:rq2b-cost-crosslang} reports the median agent steps and cost per configuration and category on both datasets. The median step count rises with category difficulty, from single-line to multi-hunk bugs, but every category on both benchmarks stays well under the 50-step cap, the highest median being 34 steps, for Defects4J MFMH.

\textbf{Under perfect localization, each individual history configuration costs no more than \nonhistory on both benchmarks, except for single-line bugs on Defects4J.} On Defects4J, only SL shows a significant cost difference (Friedman $p < 0.001$, Table~\ref{tab:rq2b-cost-crosslang}): the post-hoc pairwise Wilcoxon tests (Bonferroni $\alpha = 0.0167$) flag \fnall ($p = 0.0022$) and \fldiff ($p = 0.001$) as costing slightly more than \nonhistory, while SH, SFMH, and MFMH are comparable (all $p \geq 0.05$). On BugsInPy, no category shows a significant cost difference (all $p \geq 0.06$). A significant step-count difference appears on the hardest single-file (SFMH) BugsInPy bugs, and it favors history: the four configurations differ (Friedman $p < 0.001$) and \fldiff reaches a repair in fewer steps than \nonhistory (median 23 versus 28, $p = 0.003$). Median costs for the hardest (MFMH) repairs are also comparable or slightly lower with history (Defects4J \nonhistory \$0.029 versus \fnpair \$0.023, BugsInPy \nonhistory \$0.062 versus \fldiff \$0.060).

\begin{table}[t]
  \centering
  \caption{Effect of realistic fault localization on repair cost on DeepSeek-V3.2-Exp, measured per bug. For each configuration, on the cases it solves under both perfect and SBFL FL ($N$), we report median cost (USD) and the number of agent steps in each regime. The $p$ columns show the results of Wilcoxon signed-rank tests pairing each bug's perfect-FL value against its own SBFL value. SBFL significantly raises both cost and effort relative to perfect FL for every configuration on both benchmarks (all $p<0.05$).}
  \label{tab:rq3_sbfl_cost}
  \setlength{\tabcolsep}{5pt}
  \small
  \begin{tabular}{l l r rr r rr r}
    \toprule
                               &                 &            & \multicolumn{3}{c}{\textbf{Median Cost (USD)}} & \multicolumn{3}{c}{\textbf{Median Steps}}                                      \\
    \cmidrule(lr){4-6}\cmidrule(lr){7-9}
    \textbf{Dataset}           & \textbf{Config} & \textbf{N} & Perf.                                          & SBFL                                      & $p$      & Perf. & SBFL & $p$      \\
    \midrule
    \multirow{4}{*}{Defects4J} & \nonhistory     & 352        & 0.010                                          & 0.038                                     & $<$0.001 & 20    & 30   & $<$0.001 \\
                               & \fnall          & 362        & 0.013                                          & 0.043                                     & $<$0.001 & 20    & 28   & $<$0.001 \\
                               & \fnpair         & 369        & 0.011                                          & 0.039                                     & $<$0.001 & 20    & 29   & $<$0.001 \\
                               & \fldiff         & 362        & 0.013                                          & 0.037                                     & $<$0.001 & 21    & 29   & $<$0.001 \\
    \midrule
    \multirow{4}{*}{BugsInPy}  & \nonhistory     & 168        & 0.036                                          & 0.045                                     & 0.016    & 22    & 28   & $<$0.001 \\
                               & \fnall          & 171        & 0.030                                          & 0.059                                     & $<$0.001 & 20    & 25   & $<$0.001 \\
                               & \fnpair         & 180        & 0.034                                          & 0.040                                     & 0.026    & 20    & 27   & $<$0.001 \\
                               & \fldiff         & 191        & 0.040                                          & 0.049                                     & 0.040    & 22    & 27   & 0.003    \\
    \bottomrule
  \end{tabular}
\end{table}

\textbf{Under realistic SBFL, each individual history configuration still costs no more than \nonhistory, except \fnall.} Table~\ref{tab:rq3_sbfl_cost} reports a strict paired comparison along the FL axis: for each configuration we take the bugs solved under \emph{both} perfect and SBFL fault localization to compare cost and steps with a Wilcoxon signed-rank test that pairs each bug's two values. Relative to perfect FL, SBFL significantly increases both cost and effort for every configuration on both benchmarks (the $p$ columns of Table~\ref{tab:rq3_sbfl_cost}, all $p < 0.05$), because the agent must investigate ten suspicious locations rather than one known buggy line: on Defects4J the median cost more than triples (\nonhistory \$0.010 to \$0.038, 20 to 30 steps), and the increase is milder but still significant on BugsInPy. This extra cost comes from the harder localization, not from history. A Friedman test across configurations on cost under SBFL is significant overall (Defects4J $p = 0.019$, BugsInPy $p = 0.036$), but this comes from \fnall alone: the pairwise tests flag only \fnall as more expensive than \nonhistory (Defects4J SFMH, Wilcoxon $p = 0.006$), while \fnpair and \fldiff are statistically comparable to \nonhistory in every category, \fldiff costing the same or less (Defects4J \fldiff \$0.037 under SBFL versus \nonhistory's \$0.038). This matches \fnall's behavior under perfect FL (Table~\ref{tab:rq2b-cost-crosslang}, where its only significant difference is on SL): injecting every function name inflates tokens, whereas the diff- and pair-based heuristics do not. So whatever benefit \fnpair and \fldiff deliver under realistic FL (RQ2) is obtained without making history-augmented repair more expensive than \nonhistory.

\begin{table}[t]
  \centering
  \small
  \setlength{\tabcolsep}{5pt}
  \caption{Cost of the hybrid approach (the union of \fnall, \fnpair, and \fldiff) on DeepSeek-V3.2-Exp under perfect and realistic (SBFL) fault localization, with \fldiff\ alone as the reference. The \emph{brute-force} variant runs all three heuristics and keeps any test-passing patch, while the \emph{cascade} variant halts at the first pass in the fixed order \fnall, \fnpair, \fldiff. \emph{Fixed} counts repaired bugs, \emph{Runs} the mean heuristics executed per bug. Cost and steps are per-bug means over all $N$ bugs (including failed attempts), unlike the success-only medians of Table~\ref{tab:rq2b-cost-crosslang}. Under SBFL, $N$ covers the bugs whose four SBFL runs all recorded cost. In every setting, the hybrid costs significantly more than one \nonhistory run (paired Wilcoxon on per-bug cost, all $p<0.01$).}
  \label{tab:rq4-union-cost}
  \begin{tabular}{l l l r r r r}
    \toprule
    \textbf{FL}              & \textbf{Dataset ($N$)}                   & \textbf{Variant}    & \textbf{Fixed} & \textbf{Runs} & \textbf{Cost (\$)} & \textbf{Steps} \\
    \midrule
    \multirow{6}{*}{Perfect} & \multirow{3}{*}{\shortstack[l]{Defects4J                                                                                              \\ \scriptsize $N{=}854$}} & \fldiff\ only & 545 & 1.00 & 0.034 & 34.2 \\
                             &                                          & Hybrid, cascade     & 684            & 1.64          & 0.067              & 64.0           \\
                             &                                          & Hybrid, brute-force & 684            & 3.00          & 0.104              & 103.0          \\
    \cmidrule(lr){2-7}
                             & \multirow{3}{*}{\shortstack[l]{BugsInPy                                                                                               \\ \scriptsize $N{=}501$}} & \fldiff\ only & 374 & 1.00 & 0.086 & 29.0 \\
                             &                                          & Hybrid, cascade     & 463            & 1.36          & 0.118              & 42.2           \\
                             &                                          & Hybrid, brute-force & 463            & 3.00          & 0.240              & 87.3           \\
    \midrule
    \multirow{6}{*}{SBFL}    & \multirow{3}{*}{\shortstack[l]{Defects4J                                                                                              \\ \scriptsize $N{=}676$}} & \fldiff\ only & 419 & 1.00 & 0.078 & 37.3 \\
                             &                                          & Hybrid, cascade     & 577            & 1.58          & 0.145              & 62.5           \\
                             &                                          & Hybrid, brute-force & 577            & 3.00          & 0.239              & 111.4          \\
    \cmidrule(lr){2-7}
                             & \multirow{3}{*}{\shortstack[l]{BugsInPy                                                                                               \\ \scriptsize $N{=}265$}} & \fldiff\ only & 188 & 1.00 & 0.085 & 33.7 \\
                             &                                          & Hybrid, cascade     & 228            & 1.57          & 0.146              & 58.0           \\
                             &                                          & Hybrid, brute-force & 228            & 3.00          & 0.257              & 102.8          \\
    \bottomrule
  \end{tabular}
\end{table}

\textbf{The cascade fixes the same bugs as brute-force, but runs only 1.4 to 1.6 heuristics per bug instead of all three.} Table~\ref{tab:rq4-union-cost} prices both hybrid variants against the best individual heuristic, \fldiff, on per-bug means over all $N$ bugs including failed attempts, not the median-of-success basis of the earlier tables. Brute-force always runs all three heuristics. The cascade stops at the first heuristic that passes, and because \fnall, run first, already fixes 530 of 854 Defects4J and 377 of 501 BugsInPy bugs, most bugs never reach the second or third heuristic. Both variants reach the identical set of 684 Defects4J and 463 BugsInPy repairs, but the cascade costs about twice \fldiff\ alone on Defects4J and 1.4 times on BugsInPy, for 139 and 89 more repairs than \fldiff.

\textbf{Under realistic SBFL, the cascade still runs about 1.6 heuristics per bug and repairs the same set of bugs as brute-force.} It averages 1.58 heuristics per bug on Defects4J and 1.57 on BugsInPy, costing about 1.9 times \fldiff\ alone on Defects4J and 1.7 times on BugsInPy, for unions of 577 and 228 repairs. In all four settings, the hybrid costs significantly more than a single \nonhistory run (paired Wilcoxon, all $p<0.01$), because it runs more than one heuristic per bug where \nonhistory runs one. The two variants fix the identical set of bugs, so the cascade's lower cost has one price: it runs the heuristics one after another and stops at the first pass, whereas brute-force's three runs are independent and can execute in parallel.

\begin{rqbox}{Summary for RQ3}
  \begin{enumerate}
    \item Each individual history configuration is as efficient as \nonhistory in step count and cost, under both perfect and realistic (SBFL) localization and on both Defects4J and BugsInPy. The only exceptions are single-line Defects4J bugs under perfect FL and \fnall\ under SBFL, which cost more.
    \item The hybrid approach is affordable: its cascade variant runs the three heuristics in order and stops at the first test-passing patch, reaching the full set of 684 Defects4J and 463 BugsInPy repairs at 1.64 and 1.36 heuristics per bug, about 2.0 and 1.4 times the cost of \fldiff\ alone (1.9 and 1.7 times under realistic SBFL).
  \end{enumerate}
\end{rqbox}

\section{Discussion}\label{sec:discussion-and-future-work}

\textbf{Performance Regressions.}
On Defects4J, we observe 32 cases where a history configuration underperforms the \nonhistory one or misses \nonhistory fixes (Figure~\ref{fig:rq1_venn_combined}). For instance, in \textit{MFMH}, the \nonhistory configuration attains 50 while \fnall, \fnpair, and \fldiff reach 47, 43, and 48, respectively, even though each contributes 12 to 13 unique fixes that the \nonhistory does not solve. This pattern aligns with LLM evidence that more context can hurt: models are easily distracted by irrelevant additions \citep{shi2023large}, and performance often declines when relevant facts sit in the middle or deeper parts of long prompts \citep{levy2024same}. In our setting, extra historical snippets can dilute salient context and reduce Plausible@1 on some bugs that don't need historical context.

\smallskip\noindent\textbf{Heuristics Complementarity.}
A key finding from RQ1 (Figure~\ref{fig:rq1_venn_combined}) is that the three historical heuristics are complementary, not redundant, with each repairing a distinct set of bugs. This suggests that different bug types are susceptible to different forms of historical context: \fldiff supplies fine-grained edit patterns, \fnpair the before/after semantics of function evolution, and \fnall co-evolutionary patterns across functions. The strongest benefits appear for SFMH bugs, which involve multiple related changes within a single file, where \fldiff leads on both benchmarks. Because the blame commit typically captures the entire change context for such bugs (as shown by the single-commit concentration in RQ0), all three heuristics provide highly relevant information about the bug-introducing pattern. This complementarity is what the hybrid approach exploits: RQ3 shows its cascade variant realizes the full set of the three heuristics' repairs at 1.4 to 2.0 times the cost of an individual heuristic.

\begin{table}[t]
  \centering
  \small
  \caption{Representative bugs repaired by one of the three history heuristics but not by \nonhistory, covering both benchmarks and all four bug categories. The last column describes the blame commit's change and how it guided the fix, as recorded in the saved agent trajectory.}
  \label{tab:case-studies}
  \begin{subtable}{\columnwidth}
    \centering
    \caption{Representative bugs in Defects4J}
    \begin{tabular}{p{1.7cm} l l p{8.5cm}}
      \toprule
      Bug                   & Cat. & Heuristic & The blame commit's change, and how it guided the fix                                                     \\
      \midrule
      Closure\_113          & SL   & \fnall    & Kept \texttt{require} calls in the error-level check, a rule the agent's fix preserved                   \\
      JacksonData\ bind\_62 & SH   & \fldiff   & Moved \texttt{BeanProperty} to fluent factories, pointing the agent to the delegate-constructor handling \\
      Cli\_37               & SFMH & \fnall    & Listed the unified parser's functions, showing which parser internals to edit                            \\
      Lang\_3               & SFMH & \fldiff   & Refactored the \texttt{NumberUtils} else-clause, revealing the Float-precision logic to fix              \\
      Jsoup\_60             & MFMH & \fnall    & Named the selector-evaluator functions holding the \texttt{contains} quote-handling bug                  \\
      \bottomrule
    \end{tabular}
  \end{subtable}

  \vspace{0.5em}

  \begin{subtable}{\columnwidth}
    \centering
    \caption{Representative bugs in BugsInPy}
    \begin{tabular}{p{1.7cm} l l p{8.5cm}}
      \toprule
      Bug         & Cat. & Heuristic & The blame commit's change, and how it guided the fix                                                               \\
      \midrule
      pandas\_59  & SL   & \fnpair   & Defaulted \texttt{pairwise=True} in \texttt{corr()} when \texttt{other} is \texttt{None}, which the agent restored \\
      luigi\_17   & SL   & \fnpair   & Added \texttt{prune\_on\_get\_work=True} to the local-scheduler constructor, which the agent applied               \\
      keras\_2    & SH   & \fnpair   & Showed the \texttt{l2\_normalize} axis-reduction the agent then corrected                                          \\
      ansible\_17 & SFMH & \fnall    & Flagged the condition changed by the Facts-Refresh refactor as the likely regression                               \\
      pandas\_123 & MFMH & \fnpair   & Moved the \texttt{is\_int64\_dtype} import behind the \texttt{\_validate\_dtype} fault                             \\
      \bottomrule
    \end{tabular}
  \end{subtable}

\end{table}

\smallskip\noindent\textbf{How the Agent Uses the Injected History.}
On the 194 Defects4J and 99 BugsInPy bugs that only a history configuration repairs, we read the saved agent trajectories to see how the injected blame commit is used. Table~\ref{tab:case-studies} samples these bugs across both benchmarks, all four categories, and all three heuristics, and its last column paraphrases from each trajectory how the commit shaped the fix. The most direct case is \texttt{pandas\_59} (\fnpair), where \fnpair\ injects the \texttt{corr()} function as it looked before and after the blame commit. That earlier version set \texttt{pairwise} to \texttt{True} when \texttt{other} is \texttt{None}, a default the buggy code had dropped, so the agent reasons ``in the old implementation, when \texttt{other} is None, it sets \texttt{pairwise = True}'' and reinstates it. This case, where the injected history spells out an exact value to restore, is the exception. In most sampled bugs the agent does not revert the blame commit: the injected history points to the faulty function or reveals the earlier change behind the fault, and the agent writes its own fix. For \texttt{Cli\_37} (\fnall), the commit lists the unified parser's functions, showing which internals to edit, and for \texttt{Jsoup\_60} (\fnall) it names the selector-evaluator functions that hold the bug. Which context signal helps depends on the heuristic: the names of the functions changed in the commit (\fnall), a function's before/after form (\fnpair), or the exact edit (\fldiff). Different bugs need different signals, which is why the three heuristics repair different bug sets, the complementarity reported in RQ1 (Section~\ref{subsec:RQ1}). Furthermore, because the agent reasons from the injected commit rather than emitting the developer fix it never sees, these traces are also evidence against simple memorization (Section~\ref{sec:threats-to-validity}).

\smallskip\noindent\textbf{Plausible versus Correct Patches.}
Our primary success metric is Plausible@1 (passing the full test suite), consistent with prior work~\citep{nashid2025characterizing,ehsani2025bug,silva2025repairbench}. Yet, a patch that passes all tests is not guaranteed to be a correct fix: it can pass by coincidence or by hard-coding the expected test outputs rather than repairing the underlying bug~\citep{DBLP:conf/sigsoft/SmithBGB15,DBLP:conf/issta/QiLAR15}. Since manually verifying all ${\sim}3{,}600$ plausible patches is infeasible, we manually verify a stratified random sample of 144 of them (18 per dataset and configuration), labelling each as a correct (equivalent) or incorrect fix against the developer patch.

Overall, 59.7\% are correct (95\% confidence interval 52\% to 67\%). Scoring by correctness rather than test-passing does not overturn the RQ1 history-versus-\nonhistory comparison (Table~\ref{tab:rq2-ablation-deepseek}): pooled across the three history heuristics, the correctness rate based only on the \emph{correct} patches (instead of all plausible ones) is on par with \nonhistory on Defects4J (70\% versus 72\%) and far higher on BugsInPy (57\% versus 22\%), 64\% versus 47\% across both benchmarks, and a majority of the history-only repairs are correct (12 of 20 sampled). Because the same plausible metric and LLM are applied across all systems, including both replicated baselines and the history-versus-\nonhistory comparison, the evaluation stays fair. More broadly, as agent-based APR systems generate plausible patches at unprecedented scale, automated patch-correctness validation at scale remains a pressing open problem for the community.

\section{Threats to Validity}\label{sec:threats-to-validity}

\textbf{Internal Validity.}
DeepSeek-V3.2-Exp is available as open weights, but the pretraining corpus is not disclosed to our knowledge. As a result, we cannot verify if the specific projects or bugs in Defects4J are included in the pretraining data. However, our evaluation focuses on the relative effectiveness improvement of different historical heuristics, rather than their absolute performance. For instance, if \appname-\fldiff fixes a bug that \appname-\nonhistory does not fix, this improvement is attributed to the benefit of historical context, regardless of whether the LLM has seen the bug before.

Two further results are hard to explain by memorization alone. First, because all configurations share the same LLM, prompt scaffold, and temperature 0, the bugs that each history heuristic uniquely repairs (106 to 111 on Defects4J and 59 to 67 on BugsInPy) are attributable to the injected history rather than to memorized fixes that would be equally available to \nonhistory. Second, the complementarity is statistically significant: the hybrid approach significantly repairs bugs the \nonhistory baseline cannot (McNemar's test, $p<0.001$, Table~\ref{tab:rq2-ablation-deepseek}), and this holds under realistic SBFL as well (Table~\ref{tab:rq4-ablation}), where the retrieved commits no longer align with any developer-patch signal. To examine whether history only helps the model recall a memorized fix, Section~\ref{sec:discussion-and-future-work} reads the agent trajectories of these bugs (Table~\ref{tab:case-studies}), which show the agent reasoning from the injected blame commit rather than reciting a fix.

Our RQ1 evaluation assumes perfect fault localization, a common setup in LLM-based APR studies~\citep{yang2024sweagent,fan2023automated,bouzenia2024repairagent,xia2022less,nashid2025characterizing,ehsani2025bug}. To mitigate this threat, RQ2 evaluates \appname under realistic SBFL with GZoltar and FauxPy. Under realistic localization the complementarity remains significant (the hybrid approach still beats \nonhistory on both benchmarks), and the best history configuration remains numerically ahead, so the perfect-FL results are not merely an artifact of idealized localization.

Blameless bugs obtain history only through the nearest-line fallback (Section~\ref{subsubsec:historical-extractor}), which may retrieve a commit unrelated to the bug. On these bugs alone, \fldiff on DeepSeek repairs 174 of the 246 blameless Defects4J bugs against 147 for \nonhistory, a significant difference (McNemar's exact test, $p=0.001$), showing that the history retrieved by the fallback contributes repairs. On the 63 blameless BugsInPy bugs, the difference is not significant. Alternative fallbacks, such as blaming the enclosing method or the recent commits that touch the file, may retrieve more relevant history, and we leave their comparison to future work.


\smallskip\noindent\textbf{External Validity.}
Our study evaluates on Defects4J (854 Java bugs) and BugsInPy (501 Python bugs). While this covers two languages and 34 projects, results may not transfer to other ecosystems where project size, test adequacy, and repository practices differ. The history profiles already differ between our two benchmarks (e.g., 99.5\% single-commit in Defects4J versus 58.4\% in BugsInPy), suggesting that history-aware techniques may need adaptation per ecosystem. We expose a dataset-agnostic interface behind the history extractor and release all code and scripts for replication. Evaluating on repository-level benchmarks such as SWE-bench~\citep{jimenez2023swebench} is future work.

\appname is instantiated with one agent loop, a narrow bash tool set, specific prompts, and a single LLM with fixed step, cost, and time guards. Different agents, models, or decoding policies may change absolute numbers. API-level determinism does not guarantee identical runs. To reduce this threat, we use DeepSeek-V3.2-Exp and replicate all baselines on the same LLM.

%
%
%
%

\section{Conclusion and Future Work}\label{sec:conclusion}

This paper presented \appname, a history-aware agentic approach that injects blame-derived repository context into an observe-act-verify loop to repair single- and multi-hunk bugs. A preliminary study on 854 Defects4J (Java) and 501 BugsInPy (Python) bugs shows that blame commit history is widely available (71.1\% and 87.4\% blameable, respectively), with Defects4J blame highly concentrated in a single commit (99.5\%) while BugsInPy shows more scattered blame (58.4\% single-commit).

Using the same LLM for all experiments including replicated baselines, we find that under perfect fault localization the net benefit of history is modest (+4.4\% on Defects4J, +1.1\% on BugsInPy), but the three history heuristics are complementary: each repairs a different subset of bugs, so combining them into a hybrid approach repairs significantly more bugs than non-history. The gain from history generalizes across all three base models and is larger on the two weaker ones, and the complementarity remains significant under realistic SBFL. History does not inflate agent cost on either benchmark. On the same LLM, \appname outperforms the RepairAgent (+56.6\%) and BIRCH-feedback (+47.1\%) baselines, showing it is a strong repair agent independent of history.

Three promising directions for future work emerge from our findings. First, the complementarity of the three heuristics shows that repository history is rich: each heuristic repairs bugs the others miss, so no individual heuristic uses all of it. Future work should design more diverse heuristics over richer historical context, such as commit messages as a standalone heuristic, issue and pull-request discussions, and co-change relations, and design approaches to combine them per bug. Such an approach could also adapt the history during repair, for example by filtering it for relevance or by dropping it when the first repair steps show no progress, and reach the bugs repaired by the hybrid approach in a single run. Second, \appname currently uses an LLM judge to select a single blame commit. For BugsInPy bugs where 41.6\% have multiple blame commits, combining context from several blame commits could further improve repair effectiveness. Third, evolving \appname into a multi-agent architecture~\citep{rafi2024multi,DBLP:conf/icse/LinKC25,lee2024unified} with specialized History, Fixer, and Reviewer agents could improve repair of complex MFMH bugs where a single agent's context window is easily overwhelmed.


\section*{Data Availability}
The replication package, including source code, datasets, evaluation scripts and system prompts, is available at \url{https://github.com/SAILResearch/HAFixAgent}.





\bibliographystyle{ACM-Reference-Format}
\bibliography{sample-base}

\appendix

\section{\appname Prompt}\label{sec:hafixagent-system-prompt}

\begin{promptbox}[lst:hafixagent-prompt]{Placeholders \texttt{\{\{...\}\}} are filled at runtime.}
  You are HAFixAgent, an expert Java debugging assistant specializing in Defects4J bug repair.

  Your response must contain exactly ONE bash code block with ONE command (or commands connected with && or ||).

  ```bash
  your_command_here
  ```

  # Environment & Tools
  You operate in a Linux environment with the buggy project checked out at `{{ repo_path }}`. You have full bash access for:

  ## File Operations
  - **Overview**: `grep -n "class \|interface \|public.*(" file.java` (show classes and methods)
  - **Targeted reading**: Read around fault locations with `sed -n` for specific line ranges
  - **Progressive context**: Start with ±10 lines, expand to ±15, ±20, ±25 as needed
  - **Edit**: Precise editing with `sed -i` (simple changes) or `head`/`tail` reconstruction (complex multi-line changes)
  - **Search**: Find code patterns with `grep`, locate files with `find`

  ## Defects4J Commands
  - **Compile**: `defects4j compile` - Initial compilation to verify setup
  - **Test**: `defects4j test -r` - Compile and run relevant/failing tests

  # Bug Fixing Methodology

  ## 1. Understand the Bug
  - Read the bug description and fault locations carefully
  - Examine failing test cases to understand expected vs actual behavior
  - View the buggy code and understand its context

  ## 2. Analyze Root Cause
  - Trace through the failing test execution path
  - Identify why the current implementation produces incorrect behavior

  ## 3. Design the Fix
  - Plan minimal changes that address the root cause
  - Consider impact on other parts of the codebase
  - Ensure the fix doesn't break existing functionality

  ## 4. Implement & Verify
  - **Simple changes**: Use `sed -i` for straightforward replacements
  - **Complex changes**: Use file reconstruction with `head`/`tail` when `sed` becomes too complex
  - Test immediately: `defects4j test -r` (compiles and tests)
  - If tests still fail: analyze error output and refine the fix
  - Repeat until all tests pass

  ## 5. Multi-Hunk Strategy
  For bugs spanning multiple locations:
  - Understand the relationship between all fault locations
  - Fix locations in logical order (dependencies first)
  - Fix all related locations before testing (they often depend on each other)
  - Verify all locations work together with `defects4j test -r`

  # Success Criteria
  - All failing tests pass: `defects4j test -r` shows no failures

  **When all tests pass, signal completion with**: `echo COMPLETE_TASK_AND_SUBMIT_FINAL_OUTPUT`

  Remember: You're not just writing code - you're debugging and fixing existing systems. Think like a detective: gather evidence, form hypotheses, test them systematically.

  {% if has_blame_info %}

  # Historical Context Available
  You have access to git blame analysis showing how this code evolved. Use this context to:
  - Understand previous changes and their rationale
  - Identify patterns in how similar bugs were fixed
  - Learn from code evolution and avoid regression
  - Recognize architectural relationships and dependencies

  Pay special attention to historical context - it often reveals the "why" behind code decisions.
  {% endif %}

\end{promptbox}

\end{document}